\newcommand{\docstyle}{0} %0=elsevier or 1=ieee
\ifnum\docstyle=0
\documentclass[final,5p,times,twocolumn]{elsarticle}
\else
\documentclass[10pt,journal,compsoc]{IEEEtran}
\fi

\usepackage{acronym}
\usepackage{graphicx}
\usepackage{tabularx}
\usepackage{enumerate}
\usepackage{color}
\usepackage{amsmath}
\usepackage{paralist}
\usepackage{threeparttable}
\usepackage{booktabs}
\usepackage{colortbl,hhline}
\usepackage{boldline}
\usepackage{colortbl}
\usepackage{array}
\usepackage{adjustbox}
\usepackage{multirow,tabu}
\usepackage{algorithm,algpseudocode}
\usepackage{tikz,pgfplots}
\usetikzlibrary{arrows}
\usepackage{pgfplotstable}
\usepgfplotslibrary{units,groupplots}
\usepackage{wrapfig}
\usepackage{varwidth,xcolor}
\usepackage{soul}
\usepackage{subcaption}
\usepackage{stfloats}
\usepackage{makecell}

\def\BibTeX{{\rm B\kern-.05em{\sc i\kern-.025em b}\kern-.08em
    T\kern-.1667em\lower.7ex\hbox{E}\kern-.125emX}}
\AtBeginDocument{\definecolor{tmlcncolor}{cmyk}{0.93,0.59,0.15,0.02}\definecolor{NavyBlue}{RGB}{0,86,125}}

\usetikzlibrary{shapes.geometric, arrows.meta, positioning,patterns}
\usepgfplotslibrary{statistics}
\usepgfplotslibrary{groupplots}

\newcommand{\highlightnewtext}{0}
\ifnum\highlightnewtext=1

\else

\fi

\newcommand{\paperkeywords}{Federated Learning, Cybersecurity, Heterogeneous Data, Unbalanced Datasets}
\newcommand{\papertitle}{Federated Learning in the Wild: A Comparative Study for Cybersecurity under Non-IID and Unbalanced Settings}

\newcommand{\mytinysize}{\fontsize{6}{7}\selectfont}
\pgfplotsset{
	compat = newest,
	tick label style={font=\sffamily\scriptsize},
	label style={font=\sffamily\scriptsize},
	legend style={font=\sffamily\mytinysize\raggedleft},
	legend cell align=left,
	grid style={dotted,gray}
}

\newcolumntype{?}{!{\vrule width 0.8pt}}

\definecolor{flad}{rgb}{0.0, 0.0, 1.0}
\definecolor{fedsbs}{rgb}{0.0, 0.75, 1.0}
\definecolor{scaffold}{rgb}{1.0, 0.0, 0.0}
\definecolor{fvg}{rgb}{0.0, 0.0, 0.0}
\definecolor{fedala}{rgb}{0.6, 0.33, 0.73}
\definecolor{fedprox}{rgb}{0.0, 0.2, 0.13}
\definecolor{dafl}{rgb}{0.93, 0.5, 0.0}

\definecolor{cb_india}{HTML}{e5af53}
\definecolor{cb_cyan}{HTML}{0173b2}
\definecolor{cb_aquamarine}{HTML}{029e73}
\definecolor{cb_lilac}{HTML}{cc78bc}
\definecolor{cb_caramel}{HTML}{C91F37}
\definecolor{cb_dandelion}{HTML}{ece133}
\definecolor{cb_yellow}{HTML}{ede132}
\definecolor{cb_blue}{HTML}{0072b2}
\definecolor{cb_red}{HTML}{d55f01}

\newlength{\Oldarrayrulewidth}

\graphicspath{{artworks/}{results/}}
\DeclareGraphicsExtensions{.pdf,.jpeg,.png}
\ifnum\docstyle=0
	\journal{Computer Networks}
\fi

\begin{document}

\title{\papertitle}

\author{Roberto Doriguzzi-Corin$^\alpha$, Silvio Cretti$^\alpha$, Petr Sabel$^\beta$, Silvio Ranise$^{\alpha,\beta}$\\
\small{$^\alpha$Cybersecurity Centre, Fondazione Bruno Kessler, Trento - Italy}\\
\small{$^\beta$University of Trento, Trento - Italy}
}

% Use them with \ac{acronym name}, \acp{acronym name} for plurals, \acl{acronym name} to extend an acronym already used
\acrodef{acl}[ACL]{Access Control List}
\acrodef{ai}[AI]{Artificial Intelligence}
\acrodef{ahd}[AHD]{Abnormal Health Detection}
\acrodef{ann}[ANN]{Artificial Neural Network}
\acrodef{api}[API]{Application Programming Interface}
\acrodef{bow}[BoW]{Bag-of-Words}
\acrodef{cnn}[CNN]{Convolutional Neural Network}
\acrodef{cpe}[CPE]{Customer Premise Equipment}
\acrodef{dl}[DL]{Deep Learning}
\acrodef{dlp}[DLP]{Data Loss/Leakage Prevention}
\acrodef{dpi}[DPI]{Deep Packet Inspection}
\acrodef{dnn}[DNN]{Deep Neural Network}
\acrodef{dns}[DNS]{Domain Name System}
\acrodef{dos}[DoS]{Denial of Service}
\acrodef{ddos}[DDoS]{Distributed Denial of Service}
\acrodef{ebpf}[eBPF]{extended Berkeley Packet Filter}
\acrodef{ewma}[EWMA]{Exponential Weighted Moving Average}
\acrodef{fl}[FL]{Federated Learning}
\acrodef{flad}[FLAD]{Federated Learning Approach to DDoS attack detection}
\acrodef{flddos}[\textsc{FLDDoS}]{Federated Learning DDoS}
\acrodef{ourtool}[\textsc{FLAD}]{adaptive Federated Learning Approach to DDoS attack detection}
\acrodef{foss}[FOSS]{Free and Open-Source Software}
\acrodef{fpr}[FPR]{False Positive Rate}
\acrodef{fvg}[\textsc{FedAvg}]{Federated Averaging}
\acrodef{gdpr}[GDPR]{General Data Protection Regulation}
\acrodef{gpu}[GPU]{Graphics Processing Unit}
\acrodef{ha}[HA]{Hardware Appliance}
\acrodef{hids}[HIDS]{Host-based Intrusion Detection System}
\acrodef{ics}[ICS]{Industrial Control System}
\acrodef{ids}[IDS]{Intrusion Detection System}
\acrodef{iid}[i.i.d.]{independent and identically distributed}
\acrodef{ilp}[ILP]{Integer Linear Programming}
\acrodef{iot}[IoT]{Internet of Things}
\acrodef{isp}[ISP]{Internet Service Provider}
\acrodef{ips}[IPS]{Intrusion Prevention System}
\acrodef{jsd}[JSD]{Jensen-Shannon Distance}
\acrodef{ldap}[LDAP]{Lightweight Directory Access Protocol}
\acrodef{llm}[LLM]{Large Language Model}
\acrodef{lstm}[LSTM]{Long Short-Term Memory}
\acrodef{lucx}[LUCX]{LUCID network traffic parser eXtended}
\acrodef{mano}[NFV MANO]{NFV Management and Orchestration}
\acrodef{mbgd}[MBGD]{Mini-Batch Gradient Descent}
\acrodef{mips}[MIPS]{Millions of Instructions Per Second}
\acrodef{ml}[ML]{Machine Learning}
\acrodef{mlp}[MLP]{Multi-Layer Perceptron}
\acrodef{mqtt}[MQTT]{Message Queuing Telemetry Transport}
\acrodef{mssql}[MSSQL]{Microsoft SQL}
\acrodef{nat}[NAT]{Network Address Translation}
\acrodef{netbios}[NetBIOS]{Network Basic Input/Output System}
\acrodef{nic}[NIC]{Network Interface Controller}
\acrodef{nids}[NIDS]{Network Intrusion Detection System}
\acrodef{nf}[NF]{Network Function}
\acrodef{nfv}[NFV]{Network Function Virtualization}
\acrodef{nsc}[NSC]{Network Service Chaining}
\acrodef{ntp}[NTP]{Network Time Protocol}
\acrodef{of}[OF]{OpenFlow}
\acrodef{ood}[o.o.d.]{out-of-distribution}
\acrodef{os}[OS]{Operating System}
\acrodef{pess}[PESS]{Progressive Embedding of Security Services}
\acrodef{pop}[PoP]{Point of Presence}
\acrodef{portmap}[Portmap]{Port Mapper}
\acrodef{ppv}[PPV]{Positive Predictive Value}
\acrodef{ps}[PS]{Port Scanner}
\acrodef{qoe}[QoE]{Quality of Experience}
\acrodef{qos}[QoS]{Quality of Service}
\acrodef{rnn}[RNN]{Recurrent Neural Network}
\acrodef{sdn}[SDN]{Software Defined networking}
\acrodef{sla}[SLA]{Service Level Agreement}
\acrodef{snf}[SNF]{Security Network Function}
\acrodef{snmp}[SNMP]{Simple Network Management Protocol}
\acrodef{ssdp}[SSDP]{Simple Service Discovery Protocol}
\acrodef{svm}[SVM]{Support Vector Machine}
\acrodef{tc}[TC]{Traffic Classifier}
\acrodef{tftp}[TFTP]{Trivial File Transfer Protocol}
\acrodef{tor}[ToR]{Top of Rack}
\acrodef{tpr}[TPR]{True Positive Rate}
\acrodef{tsp}[TSP]{Telecommunication Service Provider}
\acrodef{unb}[UNB]{University of New Brunswick}
\acrodef{vm}[VM]{Virtual Machine}
\acrodef{vne}[VNE]{Virtual Network Embedding}
\acrodef{vnep}[VNEP]{Virtual Network Embedding Problem}
\acrodef{vnf}[VNF]{Virtual Network Function}
\acrodef{vsnf}[VSNF]{Virtual Security Network Function}
\acrodef{vpn}[VPN]{Virtual Private Network}
\acrodef{xdp}[XDP]{eXpress Data Path}
\acrodef{wan}[WAN]{Wide Area Network}
\acrodef{waf}[WAF]{Web Application Firewall}

\ifnum\docstyle=0
\begin{abstract}
% The pervasive reliance on digital infrastructure has amplified both the scale and sophistication of cybersecurity threats, with \ac{ddos} attacks posing significant risks to service availability. Conventional rule-based detection mechanisms often struggle to identify emerging or stealthy attack patterns, underscoring the need for adaptive solutions. In this regard, 
\ac{ml} techniques have shown strong potential for network traffic analysis; however, their effectiveness depends on access to representative, up-to-date datasets, which is limited in cybersecurity due to privacy and data-sharing restrictions. To address this challenge, \ac{fl} has recently emerged as a novel paradigm that enables collaborative training of \ac{ml} models across multiple clients while ensuring that sensitive data remains local.

Nevertheless, \ac{fvg}, the canonical \ac{fl} algorithm, has shown poor convergence in heterogeneous environments characterised by non-\ac{iid} data distributions and unbalanced dataset sizes across clients, conditions frequently observed in cybersecurity contexts. To overcome these challenges, several alternative \ac{fl} algorithms have been developed, yet their applicability to network intrusion detection remains insufficiently explored.

This study systematically evaluates a range of \ac{fl} algorithms in the context of network intrusion detection for \acs{ddos} attacks. Using a dataset of recent network attacks, the evaluation considers detection performance, training time, and communication overhead under non-\ac{iid} settings, providing practical insights into the selection of \ac{fl} solutions for network intrusion detection.
\end{abstract}

\begin{keyword}
	\paperkeywords
\end{keyword}
\maketitle
\else
\maketitle
\begin{abstract}
% The pervasive reliance on digital infrastructure has amplified both the scale and sophistication of cybersecurity threats, with \ac{ddos} attacks posing significant risks to service availability. Conventional rule-based detection mechanisms often struggle to identify emerging or stealthy attack patterns, underscoring the need for adaptive solutions. In this regard, 
\ac{ml} techniques have shown strong potential for network traffic analysis; however, their effectiveness depends on access to representative, up-to-date datasets, which is limited in cybersecurity due to privacy and data-sharing restrictions. To address this challenge, \ac{fl} has recently emerged as a novel paradigm that enables collaborative training of \ac{ml} models across multiple clients while ensuring that sensitive data remains local.

Nevertheless, \ac{fvg}, the canonical \ac{fl} algorithm, has shown poor convergence in heterogeneous environments characterised by non-\ac{iid} data distributions and unbalanced dataset sizes across clients, conditions frequently observed in cybersecurity contexts. To overcome these challenges, several alternative \ac{fl} algorithms have been developed, yet their applicability to network intrusion detection remains insufficiently explored.

This study systematically evaluates a range of \ac{fl} algorithms in the context of network intrusion detection for \acs{ddos} attacks. Using a dataset of recent network attacks, the evaluation considers detection performance, training time, and communication overhead under non-\ac{iid} settings, providing practical insights into the selection of \ac{fl} solutions for network intrusion detection.
\end{abstract}

\begin{IEEEkeywords}
	\paperkeywords
\end{IEEEkeywords}
\fi

\section{Introduction}

As reliance on digital infrastructure grows, so does the scale and complexity of cybersecurity threats.
Attacks such as data breaches, \ac{ddos} incidents, phishing campaigns, and financial fraud are becoming increasingly sophisticated, leveraging the expanded attack surface that now surrounds every connected device and user.
With cybercriminals constantly evolving their techniques, traditional rule-based security systems often suffer in identifying and mitigating novel or stealthy attack patterns.

To address these challenges, cybersecurity experts must engage in continuous research and threat analysis. This includes studying detailed attack behaviours, developing detection signatures, and maintaining extensive documentation to support the identification and mitigation of emerging threats. In addition, real-time monitoring of modern infrastructures is essential for detecting anomalies that may indicate malicious activity. However, such tasks are complex and resource-intensive, which has driven significant interest in automated solutions to improve efficiency. In this context, \acf{ml} has shown strong effectiveness across diverse application domains, including autonomous vehicles \cite{ahmad2024comprehensive}, healthcare \cite{rajesh2024threat}, \acp{ics} \cite{xue2024real, abdelaty2019aads} and maritime systems \cite{elsisi2024drone} among others, making it a promising tool for cybersecurity. Nevertheless, its adoption poses significant challenges, particularly due to strict privacy concerns and data-sharing limitations.

\acf{fl} \cite{mcmahan2017communication} has emerged as a promising approach for collaborative model training while preserving data privacy. In this framework, sensitive information remains under the control of the data owners (also called ``clients'' in the \ac{fl} context) and is not shared with other parties.  Instead, clients iteratively train on their private datasets and exchange only model updates with a central server. These updates are then aggregated to create a global model, which is progressively refined over multiple communication rounds. The ultimate goal is to enable clients to learn a common \ac{ml} model that benefits from the diversity and scale of distributed data, all while respecting the confidentiality of each client's data.

Despite its potential, \ac{fl} faces significant challenges related to model performance and training convergence. These issues are primarily a result of the heterogeneous and diverse nature of client data. In real-world scenarios, data across devices is often non-\acf{iid}, meaning local datasets can have significant differences in feature distributions, label proportions, or sampling biases (in contrast to centralised learning, where data is assumed to be \ac{iid}, meaning it's all from the same distribution). For instance, in a security application, some clients might primarily encounter specific types of attacks, while others see entirely different patterns. 

Regarding training convergence, the heterogeneity of client data can lead to model drift, where local models diverge as they adapt to their unique datasets, making it harder for the global aggregation step to converge to an optimal solution. In terms of performance, poor generalisation may arise when the global model becomes overly tailored to participants contributing large volumes of data, while those providing only limited samples are underrepresented. This imbalance introduces a bias toward data-rich clients and further undermines generalisation across the federation.

These problems also arise in the cybersecurity domain, where clients may observe highly diverse traffic patterns and feature distributions for both benign and malicious activities. For example, some clients may predominantly face specific types of attacks or variations of normal traffic, while others may record entirely different behaviours. This difference in feature distributions not only exacerbates the non-\ac{iid} nature of the data but also makes it challenging for the global model to learn a representation that generalises well across all participants. In addition, the rarity of certain attack types can lead to severe class imbalance, further complicating the training process and increasing the risk that the global model underperforms on less frequently observed threats.

To address the problems related to model performance and training convergence, researchers have developed various techniques, starting from \ac{fvg}, the canonical \ac{fl} algorithm \cite{mcmahan2017communication}, and subsequently extending it to address specific challenges. For instance, FedProx \cite{fedprox} introduces a proximal regularization term that constrains local models from diverging from the global model, directly tackling the model drift issue. SCAFFOLD \cite{scaffold} mitigates model drift by using control variates, where the server sends a correction value to each client to align their updates with the global objective before aggregation. Other approaches, such as FedALA \cite{fedala}, enhance global model generalisation on client-specific data through weighted aggregation of the local and global models. In the security domain, where data diversity is a major issue, methods like FedSBS \cite{fedsbs}, DAFL \cite{dafl}, and FLAD \cite{flad,flad-netsoft} employ alternative client selection mechanisms to ensure that a more diverse and representative group of clients participates in each training round. 
%This directly addresses the model performance problem and also helps mitigate client non-participation by ensuring a more robust and reliable training process. 
Despite these advancements, there's still no single consensus on the most effective approach, highlighting a critical gap in current research.

This work reviews recent studies that implement various \ac{fl} variants and evaluates their applicability in the context of cybersecurity network traffic analysis, specifically focusing on \ac{ddos} detection.
While some of the algorithms were originally designed for tasks unrelated to cybersecurity, our primary objective is to answer the following research question: \\
\textit{How do different strategies for improving the \ac{fl} process, including informed client selection, optimised aggregation, and enhanced local training, affect detection performance, training duration, and communication cost under heterogeneous and unbalanced client data distributions?}

Our evaluation focuses on key metrics such as training time, communication overhead, and final model accuracy, enabling a comparative analysis to assess and rank the different algorithms.
The temporal dimension of learning is especially critical in this context, as \acp{nids} must rapidly incorporate updates related to newly observed attack profiles in order to remain effective against evolving threats.
To ensure reproducibility, we configured a testbed based on the Flower framework \cite{beutel2020flower,flowerweb}, which allows for the implementation of \ac{fl} algorithms and the management of communication between clients and the server.
%This configuration mimics the deployment of FL systems in a real-world scenario where clients operate on separate devices or infrastructures.
As a data source, we utilise recent benchmark datasets that includes both benign traffic and $15$ distinct types of \ac{ddos} attack patterns \cite{unb-datasets}.

To evaluate the algorithms, we consider a federated setting characterised by severe client-level heterogeneity. Each client contains benign traffic and one specific attack type, while the amount of data varies by more than four orders of magnitude across clients. Consequently, although each local dataset is approximately balanced between benign and malicious samples, the federation is highly unbalanced with respect to both client dataset size and attack-type representation. This is called a \textit{pathological setting} \cite{mcmahan2017communication} and it represents a worst-case scenario adopted to reflect the heterogeneity commonly observed in practical cybersecurity applications, where clients may have access to very different data distributions.

In summary, the contributions of this work are the following:
\begin{itemize}
    \item \textbf{Review and adaptation of state-of-the-art \ac{fl} methods}: we review a set of \ac{fl} algorithms, including methods originally designed for other domains, and adapt them to the requirements of \acp{nids} for DDoS detection.

    \item \textbf{Multi-dimensional comparative evaluation}: we compare the considered \ac{fl} algorithms in terms of detection performance, model generalisation, training time, and communication overhead, providing a comprehensive assessment of their practical suitability for cybersecurity applications.

    \item \textbf{Analysis under heterogeneous and unbalanced conditions}: we investigate how these algorithms cope with a challenging non-\ac{iid} setting in which clients observe unique attack profiles and different data volumes, highlighting the effectiveness and limitations of different approaches to client selection, model aggregation, and local training.
    \item \textbf{Open-source implementation and reproducibility}: we provide an open-source implementation of the evaluated algorithms and the experimental setup (dataset included), enabling other researchers to reproduce our results and build upon our work \cite{fl-comparison-github}, \cite{fl-comparison-dataset}.
\end{itemize}

The remainder of this paper is organised as follows. 
Section~\ref{sec:related} analyses related work. Section~\ref{sec:problem} outlines the limitations of \ac{fvg} in heterogeneous \ac{fl} cybersecurity scenarios, while Section~\ref{sec:algorithms} reviews the main properties of the other algorithms benchmarked in this study. Section~\ref{sec:settings} provides a comprehensive description of the experimental setup, including the evaluation environment, the dataset and the hyperparameters used to configure the algorithms. Section~\ref{sec:results} presents the results of the comparative evaluation. Finally, Section~\ref{sec:conclusions} summarises the findings and concludes the study.
\section{Related work}\label{sec:related}

Although a few studies have compared \ac{fl} algorithms in the computer vision domain, our work provides additional insights by focusing on a security-specific context. Most existing research evaluates new algorithms against standard baselines, such as \ac{fvg}, with some studies extending the comparison to a few additional methods \cite{data-silos,not-all-equal}. In contrast, we broaden this scope by benchmarking a wider set of \ac{fl} algorithms within the \ac{nids} domain, with the goal of enhancing malicious traffic detection. Notably, our evaluation also includes methods specifically designed for NIDS, such as DAFL \cite{dafl}, FedSBS \cite{fedsbs}, and FLAD \cite{flad}.

Recent literature often uses the final model accuracy after a fixed number of communication rounds as the primary performance metric \cite{intrusion-evaluation} \cite{comparative-analysis} \cite{fednids} \cite{distributed-malicious}.
However, as discussed in \cite{not-all-equal}, relying solely on final accuracy can be misleading, especially under highly imbalanced data distributions where some clients have significantly more training data than others.
To address this limitation, we also report the total training duration as a measure of computational demand, following the approaches in \cite{not-all-equal} \cite{fedeval} \cite{iot-privacy}.
We also evaluate the communication overhead introduced by each method, an important metric in resource-constrained environments like Internet of Things (IoT), as emphasised by \cite{fedeval} \cite{not-all-equal} \cite{iot-privacy}.

Because our work targets lightweight models suitable for \ac{fl} environments, we restrict training to CPU-only resources, avoiding the need for GPU support.
This constraint contrasts with works such as \cite{not-all-equal} and \cite{fednids}, which rely on GPU-accelerated training for larger models.

On the server side, optimisation is particularly important in \ac{fl} \cite{not-all-equal} \cite{fedeval}, with strategies related to participant selection and workload distribution being the key to improving efficiency.
Client-side optimisation can introduce additional computational overhead, but there is a practical limit to how many computations can be offloaded to clients to enhance local training \cite{not-all-equal}.

Most prior works \cite{heterogeneity-matters} \cite{fedala} \cite{data-silos} \cite{comparative-analysis} and benchmark frameworks \cite{fedeval} assess \ac{fl} algorithm performance using datasets like CIFAR-10 or MNIST, which differ substantially from NIDS data in both structure and semantics.
Other studies \cite{fl-evaluation} distribute equal-sized datasets with differing labels among clients. Instead, we vary both label distribution and dataset size across clients.

%In contrast to prior studies, our work conducts an evaluation of a broader set of \ac{fl} algorithms within a more realistic experimental environment. Existing research commonly simulates clients by executing multiple processes on a single machine \cite{fl-evaluation} \cite{fednids}. This practice, however, has been criticised by the community \cite{fedeval}, with recommendations favouring the use of Docker containers as a more reliable alternative. With this insight, we simulate each client as an independent \ac{vm}, uniformly provisioned with an equal number of CPU cores and memory resources. This design yields a testing environment that more closely approximates real-world conditions, thereby enabling performance assessments that are both more accurate and generalizable.

While there is a growing body of research dedicated to assessing the privacy implications of \ac{fl} methods \cite{fedeval,sorbera2025adaptive}, we deliberately exclude this dimension from our study, as our primary objective is to analyse performance and efficiency rather than privacy-preserving guarantees. Similarly, although several studies provide a baseline comparison with centralised training approaches \cite{distributed-malicious}, we maintain a clear focus on decentralised learning. Our analysis is therefore entirely centered on benchmarking a diverse set of \ac{fl} algorithms with respect to their capacity to produce a global model that generalises effectively across heterogeneous client datasets. To this end, we adopt the so-called pathological settings \cite{mcmahan2017communication}, a highly challenging scenario in which each attack type is assigned to a single participant. This setup allows us to stress-test the algorithms under extreme non-\ac{iid} conditions, with the aim of highlighting their relative strengths and limitations in network intrusion detection under highly heterogeneous data distributions.
\section{Limitations of \acl{fvg}}\label{sec:problem}
\begin{figure}[t!]
    \flushright
    % trim [trim={left bottom right top},clip]
    \includegraphics[trim=8.5cm 10cm 18.5cm 8.5cm, clip,width=\linewidth]{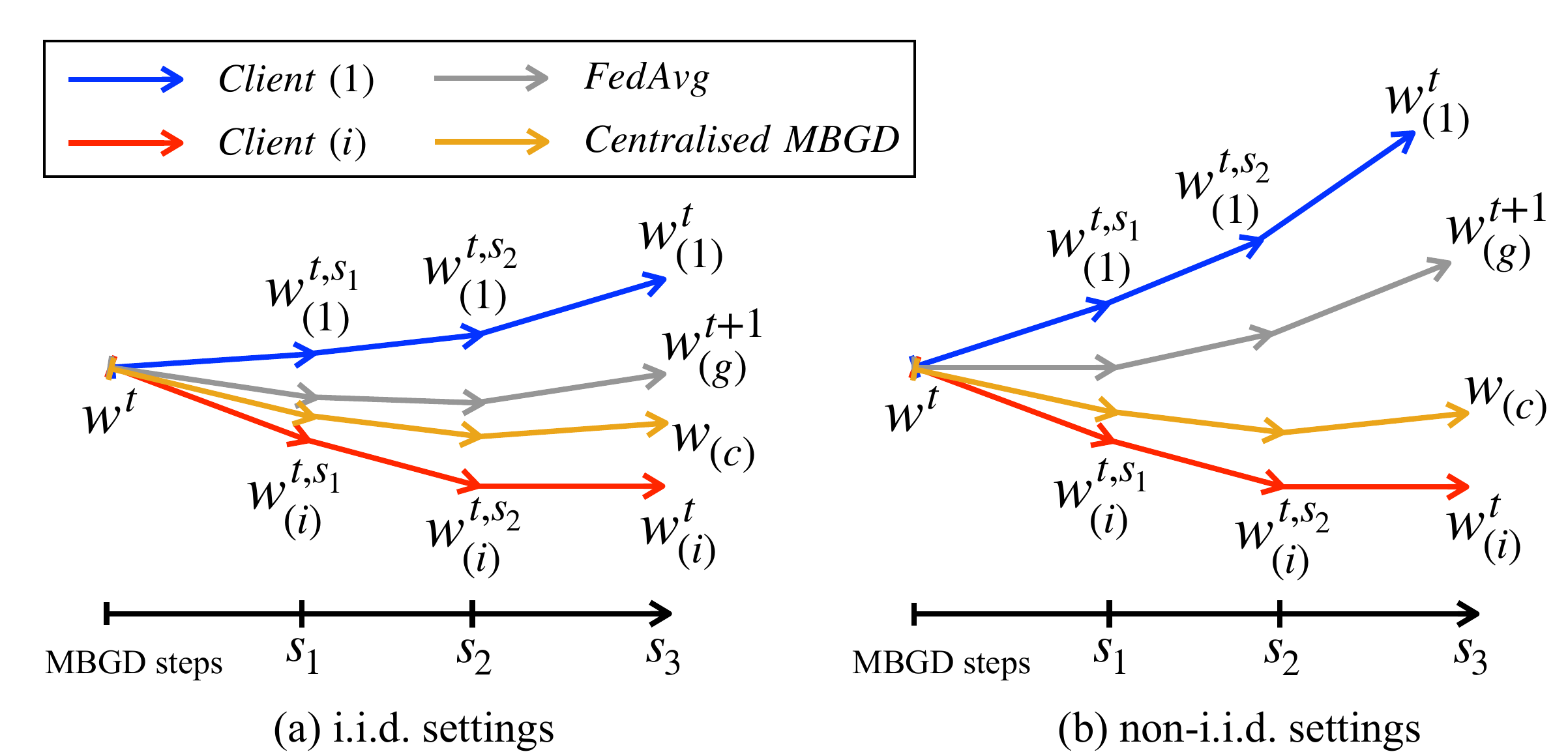}
    \caption{Weight divergence of \ac{fvg} in non-\ac{iid} settings.}
        \label{fig:iid-non-iid}
\end{figure}

Recent research \cite{wang2020optimizing,lu2024federated} has underscored the limitations of \ac{fvg} when applied to non-\ac{iid} and unbalanced datasets distributed across participating clients. In particular, the presence of non-\ac{iid} data leads to significant divergence in the local model updates, as illustrated in Figure \ref{fig:iid-non-iid}, which contrasts the \ac{iid} and non-\ac{iid} scenarios. Under non-\ac{iid} conditions, the local training performed by clients (e.g., Client $1$ and Client $i$) introduces a pronounced drift in the model parameters, causing the aggregated global model weights $w_{(g)}$ to deviate substantially from the optimal weights $w_{(c)}$ that would have been obtained under centralised training.

With \ac{fvg}, each client independently minimises its local loss function $L$ (formulated in Equation (\ref{eq:loss})) through several steps of \ac{mbgd}, starting from the shared global model $\mathbf{w}^t$ distributed at the beginning of each communication round $t$. 
\ac{mbgd} optimises the global model parameters $\mathbf{w}^t$ by computing gradients on small subsets of a client's local data (batches). During training, the client performs several epochs, where one epoch corresponds to a complete pass over its entire local dataset, ensuring that the model parameters are updated multiple times before returning the local model to the server.

Each update moves parameters in the direction that minimises the loss function:

\begin{equation}
L(\mathbf{w}^t,B) = \frac{1}{|B|}\sum_{(\mathbf{x},y)\in B} l(y,h_{\mathbf{w}^t}(\mathbf{x}))
\label{eq:loss}
\end{equation}

where, $l$ is the individual loss computed with the shared global model $\mathbf{w}^t$ on the labelled sample $(\mathbf{x},y)$, which belongs to the batch of local training samples $B$, while $h_{\mathbf{w}^t}(\mathbf{x})$ is the prediction of the model on sample $\mathbf{x}$. The full list of symbols used in this paper is provided in Table~\ref{tab:notation}.
One step of \ac{mbgd} is defined as follows:

\begin{equation}
\mathbf{w}_{(i)}^{t,s+1} = \mathbf{w}_{(i)}^{t,s}-\eta_l\nabla L(\mathbf{w}_{(i)}^{t,s},B)
\label{eq:mbgd}
\end{equation}

where the weights at step $s$ of model $\mathbf{w}_{(i)}^{t,s}$ are updated using the gradients of the loss function $L$ (with $\mathbf{w}_{(i)}^{t,0}=\mathbf{w}^t$ the global model), scaled by the client's local learning rate $\eta_l$. As shown in Figure \ref{fig:iid-non-iid}, at the end of round $t$ of local training, each client produces a local model $\mathbf{w}_{(i)}^{t}$ which is sent to the server for aggregation.

\begin{table}[t!]
    \caption{Mathematical notation used in the paper.}
    \label{tab:notation}
    \small % text size of table content
    \centering % centre the table
    \begin{adjustbox}{max width=\linewidth}
    \begin{threeparttable}
        \begin{tabular}{l l} \toprule[\heavyrulewidth]
         \textbf{Symbol} & \textbf{Description} \\ \midrule[\heavyrulewidth]
         $\mathbf{w}^t$ & the global model at round $t$\\ \hline
         $\mathbf{w}_{(i)}^{t,s}$ & the local model of client $i$ at \ac{mbgd} step $s$ of round $t$\\\hline
         $\mathbf{w}_{(i)}^t$ & the local model of client $i$ at end of round $t$\\\hline
         $\Delta \mathbf{w}_{(i)}^t = \mathbf{w}^t - \mathbf{w}_{(i)}^t$ & model update of client $i$ at round $t$\\\hline
         $D_{(i)}$ & the local training set of the $i$-th client\\\hline
         $B\subset  D_{(i)}$ & batch of samples\\\hline
         $(\mathbf{x},y)\in D_{(i)}$ & training sample $\mathbf{x}$ and its label $y$\\\hline
         $h_{\mathbf{w}^t}(\mathbf{x})$ & prediction of the model $\mathbf{w}^t$ on sample $\mathbf{x}$\\\hline
         $L(\mathbf{w}_{(i)}^t,B)$ & loss function on mini batch $B$\\\hline
         $l(y,h_{\mathbf{w}_{(i)}^t}(\mathbf{x}))$ & loss function on individual sample $(\mathbf{x},y)$\\\hline
         $\mu_i$ & number of local training updates performed by client $i$\\\hline
         $S^t$ & subset of clients chosen at round $t$\\\hline
        % \item $d^t = \sum_{i \in S^t} |D_i|$ is the sum of the length of the local datasets of the clients chosen in the round $t$.
         $N$ & the total number of clients in the federation\\\hline
         $\eta_l$ & the local learning rate\\\hline
         $\eta_g$ & the global learning rate\\\hline
         $\lambda$ & regularisation weight\\
        \bottomrule[\heavyrulewidth]
        \end{tabular}
    \end{threeparttable}
    \end{adjustbox}
\end{table}

While this procedure is effective under the assumption of \ac{iid} data, in practical scenarios where clients possess non-\ac{iid} datasets, the optimisation trajectories of the local models can diverge substantially, despite being initialised with the same parameters. This divergence arises because each client's updates reflect the biases of its own data distribution, which may not be representative of the overall population. As a result, the aggregation step performed by \ac{fvg} produces a global model that drifts away from the optimal solution that would be obtained if the loss function were minimised centrally on a unified dataset.

Moreover, when datasets sizes are unbalanced across clients, the weighted averaging scheme of \ac{fvg}, where updates are scaled by the number of local samples, amplifies the influence of clients with larger datasets. This weighting can exacerbate the bias introduced by non-\ac{iid} distributions, effectively skewing the global model toward the statistical properties of dominant clients while underrepresenting minority distributions. Consequently, \ac{fvg} may suffer from slower convergence, reduced model generalisation, and even convergence to suboptimal local minima. In extreme cases, such as when data heterogeneity is severe, the algorithm may show slow convergence and unstable behaviour, highlighting the challenges posed by non-\ac{iid} and unbalanced data in \ac{fl}.

These limitations represent a fundamental drawback of \ac{fvg} \cite{wang2020optimizing}. While prior works have primarily studied its effects in general-purpose machine learning tasks, in this work we specifically investigate its implications in the cybersecurity domain. In particular, we focus on the problem of network intrusion detection, where heterogeneous traffic distributions and unbalanced amounts of benign and malicious network traffic across clients may critically impair the performance and reliability of \ac{fl} models.
\section{Compared algorithms}\label{sec:algorithms}
In this study, we consider seven \ac{fl} algorithms, including the standard \ac{fvg} algorithm \cite{mcmahan2017communication} and six advanced methods designed to improve training in non-\ac{iid} and unbalanced settings. Three of these algorithms were specifically proposed in the context of network intrusion detection (namely, DAFL, FedSBS and FLAD), while the others were developed for more general purposes (FedProx, SCAFFOLD and FedALA). The selection of these algorithms is motivated by the goal of evaluating how different approaches address the limitations of \ac{fvg} across three key aspects of the \ac{fl} process: client selection, local training, and model aggregation.

Here we provide a brief overview of each algorithm, highlighting their key features (summarised in Table \ref{tab:algorithms} at the end of this Section).

For the sake of clarity and consistency, we use a uniform notation to describe the methods (Table \ref{tab:notation}), although it may differ from the original papers.

\subsection{FedAvg}
% Describe standard FedAvg 
\ac{fl} was introduced in 2017 by McMahan et al. \cite{mcmahan2017communication} as a communication-efficient method for training neural networks on decentralised data. At its core lies the \ac{fvg} algorithm, which was designed to enable \ac{fl} to remain effective even in challenging settings with non-\ac{iid} and unbalanced datasets. The \ac{fl} process, common to all the algorithms considered in this study, is iterated over several communication rounds, until the global model converges to a satisfactory solution (as verified by the server on a dedicated test set).

Each round consists of four main phases:
\begin{enumerate}
    \item \textbf{Model initialisation}: the server initialises the global model, which is shared with the selected participants.
    \item \textbf{Client selection}: the central server selects a subset of participants to locally train the model.
    \item \textbf{Local training}: each selected participant trains the global model $\mathbf{w}^t$ on its own local data, performing several \ac{mbgd} steps (Equations \ref{eq:loss} and \ref{eq:mbgd}). The output of this phase is a local model $\mathbf{w}_{(i)}^t$.
    \item \textbf{Aggregation}:  all participant clients send their local updates $\mathbf{w}_{(i)}^t$ to the server for aggregation to obtain a new global model $\mathbf{w}^{t+1}$.
\end{enumerate}

At each round, the participants are randomly sampled using a uniform distribution, ensuring that each client has the same probability of being selected.
The aggregation step involves a weighted averaging of model parameters:

\begin{equation}\label{eq:fvg}
\mathbf{w}^{t+1} = \mathbf{w}^{t} - \sum_{i \in S^t} \frac{|D_{(i)}|}{\sum_{i \in S^t} |D_{(i)}|} \Delta \mathbf{w}_{(i)}^t
\end{equation}

As shown in Equation (\ref{eq:fvg}), the participants with larger dataset size $|D_{(i)}|$ are assigned greater weights.

The steps outlined above are common to all \ac{fl} algorithms, although each algorithm may implement these steps differently to address data heterogeneity among clients and enhance the convergence of the global model.
In this work, we consider the \ac{fvg} algorithm as the baseline method for comparison with the other algorithms and to highlight the improvements introduced by each of them.

\subsection{General purpose algorithms}
The three algorithms presented in this section, namely FedProx, SCAFFOLD, and FedALA, were originally developed and evaluated to address the limitations of \ac{fvg} in the context of image classification. Nevertheless, their methodologies can also be adapted to cybersecurity applications, such as collaborative training of a \ac{nids}.

\subsubsection{FedProx}
FedProx \cite{fedprox} is an extension of the \ac{fvg} algorithm that introduces a regularisation term to the local loss function, allowing for more controlled updates from clients.
The regularisation term penalises the local model updates that deviate significantly from the global model. This encourages the clients to stay close to the global model during local training.
FedProx was proposed to address the challenges of heterogeneous settings, where clients have different computing capabilities and data distributions.

The local training loss $L$ introduced in Equation (\ref{eq:loss}) is modified to include a regularisation term as follows:

%$$L(w;b) = \sum_{(x,y) \in \mathbf{b}} l(w; x; y) + \frac{\mu}{2} \|w-w^t\|^2$$

$$L(\mathbf{w}_{(i)}^{t,s},B) = \frac{1}{|B|}\sum_{(\mathbf{x},y)\in B} l(y,h_{\mathbf{w}_{(i)}^{t,s}}(\mathbf{x})) + \frac{\lambda}{2} \|\mathbf{w}_{(i)}^{t,s}-\mathbf{w}^t\|^2$$

In this formulation, $i$ denotes the client index and $\lambda$ is the regularisation weight, which controls the strength of the regularisation. At the server side, the aggregation mechanism is identical to that of \ac{fvg} (Equation (\ref{eq:fvg})). FedProx does not introduce additional communication costs. However, the incorporation of the regularisation term requires clients to perform extra local computations during training.

\subsubsection{SCAFFOLD}
SCAFFOLD \cite{scaffold} introduces a correction term to the local updates, which helps to align the local models with the global model. The main idea is to correct the local updates by using a global variate $c^t$ that is shared among all clients:
\begin{equation}
\mathbf{w}_{(i)}^{t,s+1} = \mathbf{w}_{(i)}^{t,s}-\eta_l(\nabla L(\mathbf{w}_{(i)}^{t,s})+ c^t - c^t_{(i)})
\label{eq:scaffold}
\end{equation}

where $\eta_l$ is the local learning rate and $c^t_{(i)}$ is the local variate computed by client $i$ in the previous round.

Next to the model update, the local variate $c^t_{(i)}$ is updated as the gradient of the local loss function with respect to the shared global model $\mathbf{w}^t$:
$$c^{t+1}_{(i)} = \nabla L(\mathbf{w}^t,B)$$
or as follows:
$$c^{t+1}_{(i)} = c^t_{(i)}-c^t + \frac{1}{\mu_i\eta_l}(\mathbf{w}^t - \mathbf{w}^{t}_{(i)})$$
where $\mathbf{w}^{t}_{(i)}$ is the local model of client $i$ at the end of local training at round $t$, $\eta_l$ is the local learning rate and $\mu_i$ is the number of local updates performed by client $i$ at round $t$ (computed as the number of training epochs multiplied by the number of mini batches $B$). 

On the server side, the global variate is updated by incorporating the drift of the local control variates, defined as $\Delta c_{(i)} = c^{t+1}{(i)} - c^t{(i)}$, for the random subset of clients $S_t$ selected for training in round $t$. The average of the values across these clients is used to update the global control variate:

$$c^{t+1} = c^t + \frac{1}{N} \sum_{i \in S^t} \Delta c_{(i)}$$

At the beginning of round $t+1$, the global control variate $c^{t+1}$ is communicated to the subset $S_{t+1}$ and used to correct the local training as shown in Equation (\ref{eq:scaffold}).

The aggregation of the clients' local models is performed as a weighted average, as in \ac{fvg}, and scaled by the global learning rate $\eta_g$:

$$\mathbf{w}^{t+1} = \mathbf{w}^t - \eta_{g} \sum_{i \in S^t} \frac{|D_i|}{\sum_{i \in S^t} |D^i|} \Delta \mathbf{w}^t_{(i)}$$

In our validation of the SCAFFOLD algorithm, we use the first option for computing the local variate $c^{t+1}_{(i)} = \nabla L(\mathbf{w}^t,B)$, as it provides stable training with minimal computational overhead. 
In terms of communication overhead, SCAFFOLD requires the transmission of the global variate $c^t$ to all clients at each round, plus the transmission of the local variates $c^t_{(i)}$ from each client to the server.

\subsubsection{FedALA}
FedALA (Federated Learning for Adaptive Local Aggregation) \cite{fedala} combines the parameters of the global model with those of each client's local model, assigning different weights to each parameter according to its relevance for the local optimisation objective. 

The central idea of FedALA is therefore to preserve part of the information contained in the client's local model from the previous round $(t-1)$, rather than completely overwriting it with the global model received from the server in round $t$.
Specifically, because the lower layers of a neural network capture more generic representations \cite{lecun2015deep}, while the higher layers learn task-specific features, the ALA aggregation is applied only to $p$ higher layers to preserve local information, whereas the generic knowledge from the global model is retained in the lower layers. 
This is achieved by combining the current global model $\mathbf{w}^t$ with the previous local model $\mathbf{w}^{t-1}_{(i)}$ as formulated as follows:

\begin{equation}
\mathbf{\hat{w}}^{t} = \mathbf{w}^{t-1}_{(i)} + (\mathbf{w}^t-\mathbf{w}^{t-1}_{(i)})\odot [1^{|\mathbf{w}^{t-1}_{(i)}|-p}; \theta^p_{(i)}]
\label{eq:fedala}
\end{equation}

where $\odot$ is the Hadamard product, $\theta^p_{(i)}$ denotes the aggregation weights for the higher layers, whereas $1^{|\mathbf{w}^{t-1}{(i)}|-p}$ represents the identity matrix applied to the remaining $|\mathbf{w}^{t-1}{(i)}|-p$ lower layers, ensuring they are overwritten with the corresponding weights of the global model.

The resulting model $\mathbf{\hat{w}}^{t}$ is trained with \ac{mbgd} with the weights $\theta^p_{(i)}$ frozen:

$$\mathbf{w}_{(i)}^{t,s+1} = \mathbf{\hat{w}}_{(i)}^{t,s}-\eta_{l_1}\nabla_{\mathbf{\hat{w}}_{(i)}^{t,s}} L(\mathbf{\hat{w}}_{(i)}^{t,s},B)$$

Similarly, at each round the aggregation weights $\theta^p_{(i)}$ are optimised using \ac{mbgd}, while the parameters of both the global and local models remain frozen. 

$$\theta^p_{(i)} = \theta^p_{(i)}-\eta_{l_2}\nabla_{\theta^p_{(i)}} L(\mathbf{\hat{w}}_{(i)}^{t,s},B)$$
where $\eta_{l_1}$ and $\eta_{l_2}$ are the local learning rates for the model weights and the aggregation weights, respectively.

The aggregation weights are trained until convergence during the first two rounds, and in subsequent rounds they are updated for a single epoch to reduce computational overhead. 

Although FedALA does not incur additional communication costs, it introduces extra computational overhead for the clients due to the training of the ALA parameters and the aggregation of the global and local models.

\subsection{Algorithms for network intrusion detection}
DAFL, FedSBS, and FLAD were specifically designed and evaluated to address the convergence issues of \ac{fvg} in cybersecurity scenarios, particularly in the presence of heterogeneous distributions of network traffic features across clients. 
DAFL and FedSBS were evaluated on the CIC-IDS2018 \cite{ids2018dataset} and CIC-IDS2017 \cite{ids2017dataset} datasets of network traffic, which include Brute Force, DoS/DDoS, Infiltration, Botnet, and SQL Injection attacks. FLAD was evaluated on the CIC-DDoS2019 a dataset containing $13$ types of \ac{ddos} attacks \cite{ddos2019dataset}.
Each algorithm introduces modifications to the different stages of the \ac{fl} process, namely local training, client selection, and model aggregation.

\subsubsection{DAFL}
DAFL (Disparity-Aware Federated Learning) \cite{dafl} is an \ac{fl} algorithm designed for the collaborative training of \acp{nids}. Its key idea is to aggregate updates only from clients that achieve sufficiently high accuracy on their local data. The filtering is performed client-side: if a client's local model accuracy is below a predefined threshold hyperparameter $\beta$, its local model is not sent to the server. By excluding poorly performing local models, DAFL aims to accelerate convergence and improve the quality of the global model.
In addition to the filtering procedure, DAFL also modifies the aggregation strategy by giving more weight to the better-performing local models as follows:

$$\mathbf{w}^{t+1} = \sum_{i \in S^t} \frac{\rho_{(i)} a_{(i)}}{\sum_{j \in S^t} \rho_{(j)} a_{(j)}} \mathbf{w}_{(i)}^t\quad \text{where:}$$  
$$\rho_{(i)} = \frac{|D_{(i)}|}{\sum_{j \in S^t} |D_{(j)}|}\quad a_{(i)} = \frac{e^{A_{(i)}}}{\sum_{j \in S^t} e^{A_{(j)}}}$$

where $S^t$ is the subset of clients that sent back their local model at round $t$, $\rho_{(i)}$ and $a_{(i)}$ are the weights assigned to each client based on the size of their local dataset $|D_{(i)}|$ and their local accuracy $A_{(i)}$, respectively.

The DAFL algorithm is designed to improve the convergence speed of the global model by focusing on the clients that contribute positively to the training process. 
However, it may lead to a situation where some clients are never selected for aggregation, which can result in missing out some attack patterns present in their local datasets. 

It should be noted that DAFL may increase communication overhead, as the server must transmit the global model to all clients at every round. Furthermore, DAFL requires active clients, i.e., those whose validation accuracy exceeds the threshold $\beta$, to send their accuracy scores and the number of training samples to the server in addition to their local model parameters. Finally, since all clients must perform local training at each round before computing their validation accuracy, DAFL also introduces additional computational overhead.

\subsubsection{FedSBS}
The key aspect of FedSBS \cite{fedsbs} is a client selection strategy based on a greedy algorithm that selects clients by estimating their expected improvement in the global model performance with the Information Gain (IG) score.
The potential improvement is computed as the difference between the global and local losses, taking into account the imbalance $\varphi_i$ of dataset classes for each client:
$$I_i \gets - \ln{L(\mathbf{w}^{t})} + \varphi_i \ln{L(\mathbf{w}_{(i)}^{t})}$$
The local dataset imbalance coefficient $\varphi_i$ is computed by using the Shannon entropy of the class distribution in the local dataset:
$$
\varphi_i = 
\begin{cases}
    1 + \sum p_c \log_2(p_c), & \text{if } \ln{L(w_i^t) < 0} \\ 
    -\sum p_c \log_2(p_c), & \text{if } \ln{L(w_i^t) \geq 0} 
\end{cases}
$$
where $p_c$ is the proportion of class $c$ in client's $i$ dataset. The higher the value of $I_i$, the more likely the client $i$ is to be selected for the next round.
% TODO: in the code the phi is const, but in paper no -> bug 

To promote the participation of all clients, the method implements a so-called $\epsilon$-greedy policy, which selects a participant for the next round using the IG score with a probability of $1-\epsilon$, or in a random fashion with probability $\epsilon$, with $0<\epsilon<1$.
In addition, FedSBS rejects the clients that are chosen too frequently.
This is achieved by computing the Boltzmann distribution to determine the likelihood of a participant being selected for the next \ac{fl} round. Specifically, once a participant has been selected, the probability of its exclusion from subsequent selections increases with each additional selection. This approach reduces the over-participation of certain clients, thereby producing a less biased final model.

Clients of FedSBS apply a regularisation term $r(\mathbf{w}_{(i)}^{t,s})$ to the local loss function at each \ac{mbgd} step $s$:

$$L(\mathbf{w}_{(i)}^{t,s},B) = \frac{1}{|B|}\sum_{(\mathbf{x},y)\in B} l(y,h_{\mathbf{w}_{(i)}^{t,s}}(\mathbf{x})) + \lambda r(\mathbf{w}_{(i)}^{t,s})$$

where the regularisation term $r(\mathbf{w}_{(i)}^{t,s}) = \mathbf{w}_{(i)}^{t,s}+\mathbf{w}^t-\mathbf{w}^{t-1}$ incorporates the gradient information of the global model $\mathbf{w}^t-\mathbf{w}^{t-1}$ between two consecutive rounds, and $\lambda$ controls the strength of the regularisation.

The gradient information is also used by the server to implement a momentum-based method for the aggregation of clients' updates:

$$\mathbf{w}^{t+1} =  \sum_{i \in S^t} \frac{|D_{(i)}|}{\sum_{i \in S^t} |D_{(i)}|} (\mathbf{w}^{t}-\eta_g(\mathbf{w}_{(i)}^t-r(\mathbf{w}_{(i)}^{t})))$$

The momentum term serves to preserve knowledge accumulated from previous aggregation rounds, ensuring that valuable information from earlier updates is not lost. By doing so, it enhances the stability of the model during the \ac{fl} process.

FedSBS implements a mixed greedy/random client selection approach. Similar to \ac{fvg}, the number of clients selected per round can be set through a specific hyperparameter (denoted as $s$ in the FedSBS paper). Consequently, the only additional network overhead compared to \ac{fvg} arises from transmitting the clients' IG scores, which the server uses to determine the next round's participants.

\begin{table*}[t!]
    \caption{Main features of the \ac{fl} algorithms evaluated in this work. Differences with \ac{fvg} are highlighted in bold.}
    \label{tab:algorithms}
    \centering % centre the table
    \begin{threeparttable}
    \begin{tabular}{l p{40mm} p{25mm} p{35mm} p{40mm}} \toprule[\heavyrulewidth]
        \textbf{Method}  &  \textbf{Local Training} & \textbf{Client Selection} & \textbf{Aggregation}  & \textbf{Communication}\\ \midrule[\heavyrulewidth]
        \textbf{\acs{fvg}}   & \acs{mbgd} & Random subset & Weighted Averaging & Global model and client updates \\ \hline
        \textbf{FedProx}   & \acs{mbgd} with \textbf{regularisation} & Random subset & Weighted Averaging & Global model and client updates \\ \hline
        \textbf{SCAFFOLD}   & \acs{mbgd} with \textbf{control variates} & Random subset & Weighted average with \textbf{global learning rate} & Global model, client updates \newline and \textbf{Control variates}\\ \hline
        \textbf{FedALA}  & \acs{mbgd} on \textbf{aggregated local and global models};\newline \textbf{\ac{mbgd} on aggregation weights} & Random subset & Weighted averaging & Global model and client updates \\ \hline
        \textbf{DAFL}   & \acs{mbgd};\newline \textbf{Model validation} & \textbf{Accuracy threshold}\newline \textbf{on client side} & \textbf{Customised weighted averaging} based on number of training samples and local accuracy & Global model, client updates \newline and \textbf{local accuracy}\\ \hline
        \textbf{FedSBS}   & \acs{mbgd} with \textbf{regularisation} & \textbf{Information Gain client scoring} and random & Weighted averaging with \textbf{momentum} & Global model, client updates and \textbf{loss values}\\ \hline
        \textbf{FLAD}  & \acs{mbgd} with \textbf{configurable}\newline \textbf{epochs and steps};\newline \textbf{Model validation} & \textbf{Accuracy-based} & \textbf{Arithmetic mean} & Global model, client updates, \textbf{local accuracy and number of epochs and steps}\\
        \bottomrule[\heavyrulewidth]
    \end{tabular}
    \end{threeparttable}
\end{table*}

\subsubsection{FLAD}
FLAD (adaptive Federated Learning Approach to DDoS attack detection) \cite{flad} implements a custom client selection mechanism, with the aim to overcome the limitations of \ac{fvg} on unbalanced and heterogeneous \ac{ddos} attack data across the clients.
The method is designed to improve the performance of the global model by focusing on clients that are underperforming in terms of accuracy. 
This is achieved by adapting the number of local training steps performed by each client based on their local accuracy scores. Such accuracy scores are computed by the clients at each round by evaluating the global model on their local validation set and are communicated to the server. With this information, the server determines each client's local training steps for the next round.

This approach aims to assign more training to the clients with lower accuracy, so they can get closer to the optimal performance, while clients with high accuracy receive less training steps, possibly zero, because their performance is already close to the maximum.

The communication of the validation accuracy scores is essential for the FLAD algorithm, as it allows the server to determine which clients need more training steps, which ones can be skipped. The average accuracy score across clients is used as a metric to evaluate the overall performance of the global model and to determine when to stop the training process (early-stopping with \textit{patience}). However, along with the communication of the number of training steps from the server to the clients, it produces additional communication overhead compared to the standard \ac{fvg} algorithm.

The computation of accuracy scores on the local validation set introduces additional computational overhead for the clients. Unlike \ac{fvg}, where the local training load is constant across clients, the load in this case is dynamically adjusted at each round based on the client's local accuracy. As a result, the computational load may vary significantly between clients, depending on their individual performance.
\section{Experimental setting}\label{sec:settings}
The experimental environment is deployed on a server-class machine equipped with two 8-core Intel Xeon Silver 4110 @ 2.1 GHz CPUs and 64 GB of RAM. 

Each \ac{fl} algorithm has been implemented as a Flower strategy. Flower~\cite{beutel2020flower,flowerweb} is a framework that provides a high-level interface for implementing \ac{fl} algorithms and managing the communication between the server and clients. The system has been deployed using the Flower Deployment Runtime on a single server machine, with the server and clients running as separate processes. Communication between the server and clients is handled through the gRPC~\cite{grpc} protocol integrated into the Flower framework, making it completely transparent to both the server and client applications.
The source code of the environment is publicly available for evaluation and testing \cite{fl-comparison-github}.

\subsection{Algorithms and hyperparameters}\label{sec:hyperparameters}
The \ac{fl} algorithms and the global \ac{ml} model are implemented in Python, using the Tensorflow framework v2.21 for the model and the \ac{fl} algorithms. 

\subsubsection{FL algorithms}
For each \ac{fl} method, the hyperparameters are set to align with the original papers and are summarised in Table \ref{tab:hyperparameters}. 
Please note that 5 methods use the same client selection ratio ($0.5$), meaning that 7 out of 15 clients are selected at each round. The only exceptions are FLAD and DAFL, which select clients based on their local accuracy scores. With FLAD the clients are selected by the server before a \ac{fl} round starts, while in DAFL all the clients are selected for training, but their participation in the aggregated model is based on their local accuracy scores.

\begin{table}[h!]
    \caption{Hyperparameters used for each algorithm.}
    \label{tab:hyperparameters}
    \centering % centre the table
    \begin{threeparttable}
        \begin{tabular}{l l} \toprule[\heavyrulewidth]
            \textbf{Algorithm} & \textbf{Hyperparameters} \\ \midrule[\heavyrulewidth]
            \multirow{3}{*}{\textbf{\ac{fvg}}} & Selection ratio: 0.5 \\
                                             & Local epochs $E$: 1\\
                                             & Batch size $B$: 50\\ 
                                             & \ac{mbgd} steps/epoch: $\frac{\text{training samples}}{B}$\\ \hline
            \multirow{4}{*}{\textbf{FedProx}} & Regularization weight $\lambda$: 1\\ 
                                              & Selection ratio: 0.5 \\
                                              & Local epochs $E$: 1\\
                                              & Batch size $B$: 50\\ 
                                             & \ac{mbgd} steps/epoch: $\frac{\text{training samples}}{B}$\\ \hline
            \multirow{4}{*}{\textbf{SCAFFOLD}} & Global learning rate $\eta_g$: 1\\ 
                                    & Selection ratio: 0.5 \\
                                    & Local epochs $E$: 1\\
                                    & Batch size $B$: 50\\ 
                                    & \ac{mbgd} steps/epoch: $\frac{\text{training samples}}{B}$\\ \hline
            \multirow{5}{*}{\textbf{FedALA}} & Higher layers $p$: 1\\
                                    & Selection ratio: 0.5 \\
                                    & Local epochs $E$: 1\\
                                    & Sample ratio: 0.8\\
                                    & Batch size $B$: 50\\ 
                                    & \ac{mbgd} steps/epoch: $\frac{\text{training samples}}{B}$\\ \hline
            \multirow{4}{*}{\textbf{DAFL}} & Threshold accuracy $\beta$: 0.75\\ 
                                    & Selection ratio: accuracy-based \\
                                    & Local epochs $E$: 1\\
                                    & Batch size $B$: 50\\ 
                                    & \ac{mbgd} steps/epoch: $\frac{\text{training samples}}{B}$\\ \hline
            \multirow{9}{*}{\textbf{FedSBS}} & Regularization weight $\lambda$: 1\\
                                    & $\varepsilon$ min: 0.1\\
                                    & Initial $\varepsilon$: 1.0\\
                                    & Temperature: 5.0\\
                                    & Beta momentum: 0.9\\
                                    & Server learning rate: 1.0\\
                                    & Selection ratio: 0.5 \\
                                    & Local epochs $E$: 1\\
                                    & Batch size $B$: 50\\ 
                                    & \ac{mbgd} steps/epoch: $\frac{\text{training samples}}{B}$\\ \hline
            \multirow{6}{*}{\textbf{FLAD}}  & Patience value: 25 rounds\\
                                            & Selection ratio: accuracy-based\\
                                            & Min local epochs $E$: 1\\
                                            & Max local epochs $E$: 5\\
                                            & Min \ac{mbgd} steps/epoch: 10\\
                                            & Max \ac{mbgd} steps/epoch: 1000\\

            \bottomrule[\heavyrulewidth]
        \end{tabular}
    \end{threeparttable}
\end{table}

As summarised in Table~\ref{tab:hyperparameters}, the number of local epochs is set to $E=1$, meaning that each client performs a single pass over its local dataset before sending the updated model weights to the server. The number of local \ac{mbgd} steps per epoch is determined by dividing the number of samples in the local dataset by the batch size ($B=50$). Alternative configurations, such as $E=20$ or $B=10$, have also been adopted in previous work (e.g., McMahan et al.~\cite{mcmahan2017communication}). However, our preliminary experiments showed that the modest improvement in accuracy obtained by increasing the number of local \ac{mbgd} steps did not justify the substantial increase in local computation, particularly for clients with large datasets.
Note that FLAD is the only method that dynamically adjusts both the number of local epochs and the batch size, resulting in different numbers of epochs and \ac{mbgd} steps across clients and communication rounds.

To allow a fair comparison, all algorithms execute the same number of rounds.
The actual number of rounds depends on the runtime of the FLAD method, as it is the only method that implements a stopping criterion based on the average clients' validation accuracy.
FLAD executes first, and its total number of rounds is used for all other algorithms.

\subsection{The Dataset}\label{sec:dataset}

The evaluation is performed with recent datasets of \ac{ddos} attacks, IDS2017 \cite{ids2017, ids2017dataset}, IDS2018 \cite{ids2017,ids2018dataset} and DDoS2019 \cite{sharafaldin2019developing,ddos2019dataset}, provided by the Canadian Institute of Cybersecurity of the University of New Brunswick. These datasets contain several days of network activity, provided in the form of packet capture files (pcap). In this work, the pcap files are processed to extract network flows using the \ac{lucx} \cite{lucx-github}, which aggregates packet-level features into flow-level representations. Each flow is represented as a two-dimensional array of shape $(10, 32)$, where $10$ is the number of packets extracted from the traffic traces for each flow and $32$ is the number of packet-level features. These features include packet length, IP and TCP flags, and other relevant properties that characterise a traffic flow. 

The aforementioned datasets include both benign traffic and several types of network attacks. The benign traffic in the dataset was generated using the B-profile \cite{sharafaldin2018towards}, which models the distribution of typical applications such as web browsing (HTTP/S), remote shell access (SSH), file transfers (FTP), and email communications (SMTP).
The malicious traffic was produced using third-party tools and consists of 15 types of \ac{ddos} attacks, 13 of which are included in the DDoS2019 dataset, while the remaining two are found in IDS2017 and IDS2018. The attacks are described as follows:
\begin{enumerate}
    \item \textbf{WebDDoS}: A web-based DDoS attack that targets web servers, aiming to exhaust resources and disrupt service availability.
    \item \textbf{LDAP}: Uses the Lightweight Directory Access Protocol (LDAP) for reflection, increasing traffic volume directed at the victim.
    \item \textbf{PortMap}: A reflection-based DDoS attack leveraging the Port Mapper service to amplify traffic and overwhelm the target.
    \item \textbf{DNS}: Uses the Domain Name System (DNS) for reflection, amplifying traffic directed at the victim.
    \item \textbf{UDPLag}: A variation of the UDP attack, specifically designed to disrupt online gaming by introducing lag through bandwidth consumption.
    \item \textbf{NTP}: Exploits the Network Time Protocol (NTP) for reflection, amplifying traffic to overwhelm the target system.
    \item \textbf{SNMP}: Exploits the Simple Network Management Protocol (SNMP) for reflection, increasing traffic volume aimed at the target.
    \item \textbf{SSDP}: Uses the Simple Service Discovery Protocol (SSDP) for reflection, amplifying traffic to flood the victim.
    \item \textbf{Syn Flood}: A SYN flood attack that sends a series of SYN requests to a target's system, consuming resources and potentially causing a denial of service.
    \item \textbf{TFTP}: Exploits the Trivial File Transfer Protocol (TFTP) for reflection, amplifying traffic to overwhelm the target system.
    \item \textbf{UDP}: A generic DDoS attack that floods the target with User Datagram Protocol packets, consuming bandwidth and resources.
    \item \textbf{NetBIOS}: Exploits the NetBIOS service for reflection, amplifying traffic to flood the target system.
    \item \textbf{MSSQL}: Targets Microsoft SQL Server by exploiting its UDP port for reflection, amplifying traffic to the victim.  
    \item  \textbf{LOIC}: A DDoS attack tool that generates a high volume of TCP, UDP, or HTTP packets with the goal of disrupting service.
    \item  \textbf{HOIC}: A DDoS attack tool that works via an application layer HTTP Flood DDoS attack, flooding a victim's server with HTTP \texttt{GET} and \texttt{POST} requests with the goal of overloading the server's request capacity. 
\end{enumerate}

As described in previous work \cite{flad}, these attacks present disjoint feature distributions and different traffic volumes, enabling the configuration of non-\ac{iid} and unbalanced \ac{fl} scenarios. Compared to previous work, this study includes two additional TCP-based attack types (LOIC and HOIC) to better represent realistic \ac{fl} scenarios  with a more balanced distribution between TCP and UDP-based attacks.

% \begin{figure}[t!]
%     \centering
%     % trim [trim={left bottom right top},clip]
%     \includegraphics[trim=0cm 3cm 46cm 0cm, clip,width=\linewidth]{artworks/testbed.pdf}
%     \caption{Overview of the experimental environment. Each client is assigned a dataset containing benign traffic and a single \ac{ddos} attack type. The number of samples refers to the overall dataset size, encompassing training, validation, and test sets.}
%     \label{fig:k8s-arch}
% \end{figure}

\begin{table}[t]
\centering
\caption{Overview of the experimental environment. Each client is assigned a dataset containing benign traffic and a single \ac{ddos} attack type.}
\label{tab:clients}
\resizebox{\linewidth}{!}{\begin{tabular}{llrrr}
\Xhline{1.1pt}
\textbf{Dataset} & \textbf{Client/Attack} & \textbf{Train} & \textbf{Validation} & \textbf{Test} \\
\Xhline{1.1pt}
\multirow{13}{*}{DDoS2019}
& 1-WebDDoS (TCP) & 323 & 37 & 40  \\
& 2-LDAP (UDP)      & 646 & 72 & 82  \\
& 3-Portmap (UDP)   & 1,296 & 145 & 161  \\
& 4-DNS (UDP)       & 2,592 & 288 & 320  \\
& 5-UDPLag (UDP)    & 5,182 & 578 & 640  \\
& 6-NTP (UDP)       & 10,368 & 1,152 & 1,280  \\
& 7-SNMP (UDP)      & 20,732 & 2,304 & 2,566  \\
& 8-SSDP (UDP)      & 41,472 & 4,608 & 5,120  \\
& 9-Syn Flood (TCP) & 82,934 & 9,222 & 10,251  \\
& 10-TFTP (UDP)      & 165,888 & 18,433 & 20,481  \\
& 11-UDP Flood (UDP) & 331,776 & 36,864 & 40,960  \\
& 12-NetBIOS (UDP)   & 663,552 & 73,728 & 81,920  \\
& 13-MSSQL (UDP)     & 1,327,108 & 147,460 & 163,841  \\
\midrule
IDS2017 & 14-LOIC (TCP) & 161,060 & 17,896 & 19,885  \\
\midrule
IDS2018 & 15-HOIC (TCP) & 1,480,599 & 164,512 & 182,791 \\
\Xhline{1.1pt}
\multicolumn{2}{l}{\textbf{Total}} & \textbf{4,105,697} & \textbf{456,299} & \textbf{507,338} \\
\Xhline{1.1pt}
\end{tabular}}
\end{table}

\subsection{Clients}
The evaluation environment consists of 15 clients and a single central server that communicates with all clients.
As in previous research \cite{flad}, the dataset is distributed among the clients in a non-\ac{iid} and unbalanced manner: each attack type is assigned to a different client, and the amount of data varies across clients as presented in Table \ref{tab:clients}. This scenario simulates an \ac{fl} environment where clients possess heterogeneous data both in terms of distribution and quantity.

Each client's dataset is balanced to contain approximately the same amount of benign and attack samples. Moreover, each dataset is split into training (90\%) and test (10\%) sets, with 10\% of the training set kept as validation data. It is worth noting that the test set is never used during the \ac{fl} process and is only employed to evaluate the final global model performance.

\subsubsection{Global model}\label{sec:global_model}

The global model is a \ac{mlp} consisting of two hidden layers with 32 neurons each, using the ReLU activation function. ReLU is a widely adopted activation function in neural networks due to its low computational cost and its ability to mitigate the vanishing gradient problem \cite{nair2010rectified}.

The input to the model is an encoded representation of a network flow that aggregates packet-level features. As described in Section \ref{sec:dataset}, a single sample is a two dimensional array of shape $(10, 32)$, where $10$ is the number packets extracted from the traffic traces for each flow and $32$ is the number of packet-level features. 
The neural network is trained to classify the network traffic either as benign or \ac{ddos}. Thereby, the output layer uses a sigmoid activation function to produce a probability score for each traffic flow being \ac{ddos}. The loss function is the binary cross-entropy, a common choice for binary classification tasks. Model parameters are optimised using \ac{mbgd} with learning rate set to $0.1$.

Model weights are randomly initialised by the server using a different seed for each test run and are communicated to all clients before starting the \ac{fl}. To ensure a fair comparison and reproducible results, the initialised \ac{mlp} model is kept the same across all \ac{fl} methods.

\section{Evaluation results}\label{sec:results}
As anticipated in the previous sections, the experiments are designed to evaluate the performance of \ac{fl} algorithms in \textit{pathological} settings with non-\ac{iid} and unbalanced data distributions across clients. We compare the performance of the algorithms in terms of network overhead, client selection, training time, model aggregation, and accuracy and analyse the impact of their customisations with respect to the standard \ac{fvg} algorithm.
With these experiments, we aim to provide a comprehensive understanding of the strengths and limitations of each algorithm in the context of cybersecurity, where the global model must generalise across diverse attack types, be updated quickly to minimise the time clients remain exposed to new attacks, and incur minimal network overhead to avoid saturating communication channels in scenarios with limited bandwidth and large model sizes.

For each metric, we perform 10 runs with different random seeds to improve the reliability of the results.

\begin{table*}[t!]
    \caption{Average network bandwidth (MB) consumed by each client across all 10 experiments.}
    \label{tab:thput}
    \centering % centre the table
    \begin{threeparttable}
    \resizebox{\textwidth}{!}{\begin{tabular}{lccccccccccccccc|c } \toprule[\heavyrulewidth]
        \textbf{Method}  & \textbf{WebDDoS} & \textbf{LDAP} & \textbf{Portmap} & \textbf{DNS} & \textbf{UDPLag} & \textbf{NTP} & \textbf{SNMP} & \textbf{SSDP} & \textbf{Syn} & \textbf{TFTP} & \textbf{UDP} & \textbf{NetBIOS} & \textbf{MSSQL} & \textbf{LOIC} & \textbf{HOIC} &\textbf{Total} \\ \midrule[\heavyrulewidth]
        \textbf{\ac{fvg}} & 6.00 & 5.89 & 5.89 & 5.89 & 5.80 & 6.01 & 6.05 & 6.02 & 6.10 & 6.13 & 6.05 & 6.25 & 6.00 & 5.73 & 6.13 & 90 \\
        \textbf{FedProx} & 6.00 & 5.89 & 5.89 & 5.89 & 5.81 & 6.02 & 6.05 & 6.03 & 6.10 & 6.13 & 6.05 & 6.25 & 6.00 & 5.73 & 6.13 & 90 \\
        \textbf{SCAFFOLD} & 11.96 & 11.74 & 11.76 & 11.74 & 11.58 & 11.99 & 12.07 & 12.01 & 12.16 & 12.23 & 12.07 & 12.47 & 11.98 & 11.43 & 12.23 & 179 \\
        \textbf{FedALA} & 6.00 & 5.89 & 5.90 & 5.89 & 5.81 & 6.02 & 6.05 & 6.03 & 6.10 & 6.14 & 6.05 & 6.25 & 6.01 & 5.73 & 6.14 & 90 \\
        \textbf{DAFL} & 12.13 & 12.83 & 12.84 & 12.84 & 12.84 & 12.84 & 12.84 & 12.84 & 12.84 & 12.84 & 12.84 & 12.84 & 12.84 & 12.82 & 12.83 & 192 \\
        \textbf{FedSBS} & 5.97 & 6.16 & 5.94 & 5.80 & 5.97 & 5.91 & 6.23 & 6.01 & 6.03 & 6.06 & 6.13 & 5.79 & 6.14 & 5.93 & 5.94 & 90 \\
        \textbf{FLAD} & 16.90 & 17.54 & 16.19 & 7.86 & 17.20 & 16.63 & 7.32 & 6.54 & 8.00 & 6.56 & 7.04 & 13.65 & 7.23 & 17.91 & 11.68 & 178 \\
        \bottomrule[\heavyrulewidth]
    \end{tabular}}
    \end{threeparttable}
\end{table*}

\subsection{Network overhead}
Network overhead is a crucial aspect of \ac{fl} algorithms, especially in environments with limited, unstable or expensive connections. In this section, we analyse the network traffic generated by each approach, focusing on the amount of data exchanged between the clients and the server during training, including global and local model parameters, as well as any additional information required to manage the \ac{fl} process. The network traffic overhead introduced by the Flower platform is excluded, as it is independent of the specific \ac{fl} algorithm and common to all evaluated approaches.
In this way, we also highlight the different approaches to client selection and model aggregation.
Our experiments are conducted with the \ac{mlp} model described in Section \ref{sec:global_model}, which consists of more than $11$K trainable parameters and has a size of around $45$ KBytes. 

%\subsubsection{Network overhead with the default model size}
Table \ref{tab:thput} presents the experimental results about network traffic consumption, with \ac{fvg} as the baseline.
The network consumption data for FedProx, FedSBS and FedALA is similar to that of \ac{fvg}. FedProx and FedALA use the same client selection strategy and do not require any additional data exchange. FedSBS, on the other hand, requires clients to send their IG scores to the server for client selection. However, the size of these loss values is negligible compared to the size of the model parameters, resulting in a similar network overhead to \ac{fvg}.
On the other hand, the network overhead introduced by  SCAFFOLD, DAFL and FLAD differs due to their specific design choices:
\begin{itemize}
    \item \textbf{SCAFFOLD} requires exchanging arrays of control variates, which have the same number of parameters as the model itself.
    \item \textbf{DAFL} always requires all clients to train, although only those that meet a predefined accuracy threshold send their local model to the server.
    \item \textbf{FLAD} evaluates the performance of the global model on local datasets and requires clients to send their local accuracy values to the server. Consequently, the server sends the global model to all clients in every round for evaluation. %However, the number of clients that upload their local model to the server is dynamically adjusted based on their performance, reducing the network overhead as training proceeds.
\end{itemize}

%The network overhead of both DAFL and SCAFFOLD is significantly higher than that of \ac{fvg}. Although their mechanisms differ, the amount of data exchanged in each round is very similar. 
%With FedSBS, $25$ models are exchanged between server and clients: the server sends $6$ copies of the global model to the selected clients, which in turn send one local model each back to the server. In addition, the server distributes a copy of the global model to all $13$ clients for the evaluation of global and local losses on their validation sets.
With DAFL, the server always sends a copy of the global model to all $15$ clients. Only those clients whose local accuracy exceeds a predefined threshold $\beta$ return their model to the server. In theory, the number of responding clients could range from $0$ to $15$. In practice, with our setup, all the clients consistently achieve a local accuracy above the threshold ($\beta=0.75$). The only exception is the WebDDoS client, which fails to meet the accuracy requirement in around $15$ rounds on average across all the experiments.
This results in approximately $30$ model exchanges per round, more than double of the $14$ exchanges of \ac{fvg}.
%In addition, both approaches require clients to send extra information to the server: FedSBS transmits the losses, while DAFL transmits the local accuracies used for aggregation. This further explains why the network overhead of the two methods is so similar. 

The network overhead of SCAFFOLD is influenced by the exchange of control variates between clients and the server. In our implementation, the control variates are computed as the gradient of the global model on the local dataset and stored in a NumPy array with the same number of elements as the model's weights. Since a TensorFlow model also includes additional metadata (e.g., model's architecture, optimizer state, training configuration, etc.), its size is slightly larger than that of the control variate array. At each round, the server sends the global model and the global control variates to seven selected clients and receives seven local models and seven control variate arrays in return. In total, therefore, $28$ data structures are exchanged between the server and clients at each round, close to the $30$ exchanged by DAFL. However, 14 of them are control variates, whose size is smaller than a model. %Moreover, FedSBS and DAFL also require the transmission of additional information such as losses and accuracies. Overall, this explains the observed differences in network overhead among the three approaches.

As already mentioned in Section \ref{sec:algorithms}, FLAD and FedSBS implement an evaluation phase, which is necessary to compute the performance of the global model on each client's local dataset. This evaluation is crucial for both methods, as it allows them to adaptively select clients based on their local performance. However, there is a key difference between the two methods: while FedSBS always selects a constant number of clients per round (partly at random and partly based on the IG score), which compute the IG score just after training, FLAD adaptively adjust the number of clients selected in each round as training progresses (see Figure \ref{fig:flad-participants}), focusing on those clients that require more training to improve the global model's performance. This dynamic selection strategy helps FLAD to reduce network overhead caused by local updates transmitted by the clients, as it progressively excludes clients that have already achieved good performance. As shown in Table \ref{tab:thput}, FLAD significantly reduces network consumption for clients whose attack traffic is easier to learn, such as DNS, SNMP, SSDP, Syn, TFTP, UDP and MSSQL.  However, it is important to note that FLAD always requires all clients to receive the global model in each round for evaluation, which contributes to its overall network overhead.

As a final consideration, the \ac{dl} models used in these experiments are relatively small ($45$ KBytes) and introduce negligible network overhead in most scenarios. In practice, however, models can be much larger, for example, \acp{llm} such as Bidirectional Encoder Representations from Transformers (BERT), which are also applied in cybersecurity \cite{li2025securebert}. These models contain hundreds of millions of trainable parameters and can exceed 1 GByte in size. In such cases, FedProx, FedALA or FedSBS would be the preferred approaches for training a \ac{nids} in heterogeneous settings.
\begin{table*}[!t]
    \caption{Percentage of rounds in which a particular client is selected for local training (with respect to the total number of rounds). Approaches using \ac{fvg}'s random selection (FedProx, SCAFFOLD and FedALA) and FedSBS choose 7 out of 15 clients in each round, resulting in an average participation rate of about 47\%.}
    \label{tab:rounds}
    \centering % centre the table
    \begin{threeparttable}
    \resizebox{\textwidth}{!}{\begin{tabular}{lccccccccccccccc|c } 
    \toprule[\heavyrulewidth]
        \textbf{Method}  & \textbf{WebDDoS} & \textbf{LDAP} & \textbf{Portmap} & \textbf{DNS} & \textbf{UDPLag} & \textbf{NTP} & \textbf{SNMP} & \textbf{SSDP} & \textbf{Syn} & \textbf{TFTP} & \textbf{UDP} & \textbf{NetBIOS} & \textbf{MSSQL} &  \textbf{LOIC} & \textbf{HOIC} & \textbf{Average} \\ 
        \midrule[\heavyrulewidth]
        \textbf{\ac{fvg}-based} & 46.7 & 45.8 & 45.9 & 45.8 & 45.2 & 46.8 & 47.1 & 46.9 & 47.4 & 47.7 & 47.1 & 48.6 & 46.7 & 44.6 & 47.7 & 46.7 \\
        \textbf{DAFL}           & 100.0 & 100.0 & 100.0 & 100.0 & 100.0 & 100.0 & 100.0 & 100.0 & 100.0 & 100.0 & 100.0 & 100.0 & 100.0 & 100.0 & 100.0 & 100.0 \\
        \textbf{FedSBS}         & 46.4 & 47.9 & 46.2 & 45.1 & 46.4 & 45.9 & 48.4 & 46.7 & 46.9 & 47.2 & 47.7 & 45.0 & 47.7 & 46.1 & 46.2 & 46.7 \\
        \textbf{FLAD}           & 81.4 & 86.4 & 75.9 & 11.0 & 83.7 & 79.3 & 6.8 & 0.7 & 12.1 & 0.9 & 4.6 & 56.1 & 6.1 & 89.3 & 40.8 & 42.3 \\
        \bottomrule[\heavyrulewidth]
    \end{tabular}}
    \end{threeparttable}
\end{table*}

\begin{figure}[b!]
    \pgfplotstableread[col sep=comma]{results_data/flad_participants.dat}{\participants}
    \centering
    \begin{tikzpicture}
        \begin{axis}[
            % legend on two rows
            legend columns = 2,
            legend style = {at={(0.54, 0.7)}, anchor=south, inner sep=3pt, style={column sep=0.15cm}},       %vorher: at={(1.3, 1)}
            legend cell align=left,
            xmin = 0, xmax=250, %4100
            ymin = 0, ymax = 15, %155000
            xtick distance=50,
            xtick={0,50,...,250},
            xticklabels={0,50,...,250},
            xtick pos=left,
            ytick distance=1,
            ytick={0,...,15},
            yticklabels={0,...,15},
            grid = both,
            width=1.07\linewidth,
            height=6cm,
            xlabel = {Round},
            ylabel = {Clients},
            enlargelimits=0.02
            ]

            \addlegendentry{FLAD}
            \addlegendentry{DAFL\ \ \ \ \ \ \ \ \ \ \ \ \ \ \ \ \ \ \ \ \ \ \ \ \ \ \ \ \ \ \ \ \ \ \ \ \ \ \ \ \ \ }
            \addlegendentry{\makebox[10pt][l]{{\ac{fvg}, FedALA, FedProx, FedSBS, SCAFFOLD}}}
    
            % Layers DAFL FedALA FedAvg FedProx FedSBS FLAD SCAFFOLD
            \addplot[color=flad, style={thick},mark=*] table [x = {Epoch}, y = {FLAD}] {\participants};
            % \addplot[color=blue, style={thick}, mark=square*, mark size=1.8] table [x = {Epoch}, y = {FLAD2}] {\participants};
            % \addplot[color=orange, style={thick},mark=triangle*, mark size=2.5] table [x = {Epoch}, y = {FLAD3}] {\participants};
            \addplot[color=dafl,style={thick,dashed}] table [x = {Epoch}, y = {DAFL}] {\participants};
            \addplot[color=fvg, style={thick}] table [x = {Epoch}, y = {Others}] {\participants};
        \end{axis}
    \end{tikzpicture}
    \caption{Clients/round selected during the longest experiment run (250 rounds). The two horizontal lines represent the static number of clients/round selected by DAFL (15 clients) and the other methods (7 out of 15 clients).}
    \label{fig:flad-participants}
\end{figure}

\subsection{Client selection}\label{sec:results:client_selection}

A crucial aspect of \ac{fl} in heterogeneous environments is the client selection strategy. The choice of which clients participate in each training round can significantly impact the convergence speed, final model accuracy and the computational overhead on clients. Different \ac{fl} methods employ various strategies for client selection, ranging from random selection to more sophisticated approaches based on client performance or data characteristics.
To evaluate the effectiveness of each selection strategy, we report the distribution of training rounds across all clients in Table \ref{tab:rounds}.
Most of the analysed methods incorporate the \ac{fvg}'s random selection, which assigns a similar number of rounds to each client.
The analysis is more interesting for the DAFL, FedSBS, and FLAD algorithms.

DAFL performs client selection after the training procedure, forcing all clients to execute local training in every round.
This results in a long training time because the slowest client always performs the local training.
FedSBS selects clients partly at random and partly through a scoring mechanism based on the Information Gain metric. Similar to \ac{fvg}, it also enforces a maximum client participation per round ($50\%$). While its design explicitly aims to prioritise clients that maximise Information Gain, in practice the approach proves less effective: under the default configuration, the resulting client selection distribution closely resembles that of \ac{fvg}, showing near-uniform randomness.

FLAD, on the other hand, does not impose a fixed number of clients per round, but instead selects clients based on their local accuracy scores. As shown in Table \ref{tab:rounds}, clients such as WebDDoS, LDAP, Portmap, UDPLag, NTP and LOIC are selected more frequently than others. A higher number of training rounds indicates that more computation is needed to build an effective model for that client.

Figure \ref{fig:flad-participants} shows the number of selected participants per round during the longest experiment run. In the figure, we can see that DAFL selects all clients in each round, while FLAD selects a varying number of participants over time, with an average of approximately $6.42$ participants per round. All the other methods select 7 out of 15 clients, which is the result of a selection ratio of $0.5$.

% \hl{verify the following}
% Although FLAD selects fewer clients per round on average, this does not implies that the duration of the whole \ac{fl} is shorter. As discussed in Sections \ref{sec:results:time} and \ref{sec:results:local-time}, FLAD's training time can be longer than that of other methods, as it dynamically assigns training load to each client. This can lead to very long local training sessions for some clients, hence increasing the overall \ac{fl} duration.
\subsection{Federated training time}\label{sec:results:time}
The duration of \ac{fl} is critical, as timely model updates are essential in cybersecurity to maintain system performance. In this section, we analyse the duration of each method in terms of the total time taken to train the model and the time spent by each client during the training process. 
\begin{figure*}[b!]
	\begin{subfigure}[t]{0.49\textwidth} 
    \begin{tikzpicture}
    \pgfplotstableread[col sep=comma]{results_data/duration_histogram.dat}\datatable
    \pgfplotsset{boxplot/whisker range=1000}

    \begin{axis}[
        width=\textwidth,height=5cm,
        scaled y ticks=base 10:-3,
        ylabel={Time ($\times 10^3$ seconds)},
        scaled y ticks=false,
        ymin=0,
        ymax=30000,
        ytick distance=5000,
        ytick={0,5000,...,30000},
        yticklabels={0,5,...,30},
        grid=both,
        boxplot/draw direction=y,
        xtick={1,...,7},
        xticklabels={FedAvg,FedProx,SCAFFOLD,FedALA,DAFL,FedSBS,FLAD},
        % rotate labels by 45 degres
        xticklabel style={rotate=40, anchor=east},
        % style for the box bodies
        boxplot/every box/.style={
        draw=black,            % box outline color
        fill=white!30,          % box fill color (same for all)
        line width=0.8pt
        },
        % style for the median line
        boxplot/every median/.style={
        draw=blue!90!black,
        line width=1.2pt
        },
        % style for whiskers and whisker caps
        boxplot/every whisker/.style={
        draw=black,
        line width=0.9pt
        },
        boxplot/every whisker cap/.style={
        draw=black,
        line width=0.9pt
        },
        % reduce horizontal gap between boxes
        boxplot/box extend=0.5,
        enlargelimits=0.05
        ]
        % Add one boxplot per column. The loop uses the column name as y=...
        \foreach \col in {FedAvg,FedProx,SCAFFOLD,FedALA,DAFL,FedSBS,FLAD} {
        \addplot[boxplot] table[y=\col] {\datatable};
        }

    \end{axis}
    \end{tikzpicture}
    \vspace{-5mm}
    \caption{Total training time.}
    \label{fig:total_duration}
    \end{subfigure}
    \hfill
    \begin{subfigure}[t]{0.49\textwidth} 
    \begin{tikzpicture}
    \pgfplotstableread[col sep=comma]{results_data/duration_histogram_local_training.dat}\datatable
    \pgfplotsset{boxplot/whisker range=1000}

    \begin{axis}[
        width=\textwidth,height=5cm,
        scaled y ticks=base 10:-3,
        ylabel={Time ($\times 10^3$ seconds)},
        scaled y ticks=false,
        ymin=0,
        ymax=30000,
        ytick distance=5000,
        ytick={0,5000,...,30000},
        yticklabels={0,5,...,30},
        grid=both,
        boxplot/draw direction=y,
        xtick={1,...,7},
        xticklabels={FedAvg,FedProx,SCAFFOLD,FedALA,DAFL,FedSBS,FLAD},
        % rotate labels by 45 degres
        xticklabel style={rotate=40, anchor=east},
        % style for the box bodies
        boxplot/every box/.style={
        draw=black,            % box outline color
        fill=white!30,          % box fill color (same for all)
        line width=0.8pt
        },
        % style for the median line
        boxplot/every median/.style={
        draw=blue!90!black,
        line width=1.2pt
        },
        % style for whiskers and whisker caps
        boxplot/every whisker/.style={
        draw=black,
        line width=0.9pt
        },
        boxplot/every whisker cap/.style={
        draw=black,
        line width=0.9pt
        },
        % reduce horizontal gap between boxes
        boxplot/box extend=0.5,
        enlargelimits=0.05
        ]
        % Add one boxplot per column. The loop uses the column name as y=...
        \foreach \col in {FedAvg,FedProx,SCAFFOLD,FedALA,DAFL,FedSBS,FLAD} {
        \addplot[boxplot] table[y=\col] {\datatable};
        }

    \end{axis}
    \end{tikzpicture}
    \vspace{-5mm}
    \caption{Local training time: total local time of the slowest client at each round.}
    \label{fig:local_duration}
    \end{subfigure}
    \caption{Distribution of execution durations for each method, reporting the minimum and maximum observed values. The boxplots indicate the interquartile range (i.e., the range between the 25th and 75th percentiles), with the horizontal blue line denoting the average value.}
    \label{fig:duration}
\end{figure*}

In this experiment, all \ac{fl} methods are trained for the same number of rounds. As discussed in Section \ref{sec:hyperparameters}, the number of rounds is determined by FLAD, which is the only method that applies an early stopping mechanism. FLAD is executed first, and the other methods are then run for the same number of rounds to ensure fairness. 

The duration of the \ac{fl} process is shown in Figure \ref{fig:total_duration}, which presents the distribution of execution times for each method, reporting the minimum and maximum observed values, the interquartile range, and the average. The reported duration includes the overhead introduced by the Flower framework (including client orchestration, gRPC communication, and message serialization/deserialization), client-side operations (e.g., model training and evaluation), data transmission (model weights, control variates, accuracy scores, etc.), and global model aggregation.

Figure \ref{fig:local_duration} shows the contribution of the clients' local operations to the overall training time. Specifically, it reports the cumulative execution time of the slowest client at each federated learning round, as the server must wait for all selected clients before proceeding. The results indicate that local operations account for the largest share of the total training time (approximately 60\%) for all methods except FLAD. In FLAD, local training represents only about 9\% of the total execution time. Consequently, the remaining time (between 4500 and 5500 seconds) is largely dominated by the Flower framework overhead described above, including client orchestration, communication, and synchronization.

The main advantage of FLAD is its adaptive client selection mechanism, which prioritises clients whose data is harder to classify. Client selection is based on the accuracy of the global model on each client's local validation set, so more challenging clients are selected more frequently and assigned more training epochs and steps, while easier clients are selected less often. Since clients train in parallel, the duration of each round is determined by the slowest selected client, typically the one with the largest local dataset. As shown in Table \ref{tab:rounds}, some large clients such as \textit{MSSQL}, \textit{UDP}, and \textit{TFTP} are selected much less frequently than in other methods, reducing the overall duration of the \ac{fl} process.

Other relevant observations from Figure \ref{fig:duration} are:
\begin{itemize}
    \item The duration of the \ac{fl} process with \textbf{FedProx}, \textbf{SCAFFOLD}, \textbf{FedALA} and \textbf{FedSBS} aligns with that of \ac{fvg}, as they require minimal extra computation to train the clients. FedProx introduces a small overhead due to the additional regularisation term, SCAFFOLD requires the computation of control variates, which does not significantly affect the training time. FedALA requires an additional training phase to estimate the aggregation weights, which adds some overhead, but this is not significant compared to the overall training time. The execution time of FedSBS slightly exceeds that of \ac{fvg} due to the additional evaluation phase, during which the global model is systematically assessed on each client's local dataset.
    \item \textbf{DAFL} requires all clients to train in each round, which leads to a longer duration compared to \ac{fvg}. This is because the slowest client (that with the largest dataset \textit{HOIC}) determines the overall duration of each round, and with DAFL this client is always selected.
\end{itemize}

% The overall duration of the \ac{fl} process is determined by the time taken by the slowest client selected in each round to execute local operations and the time required to transmit the model weights and other data to the server.
% In our experimental setup, all the clients are running on machines with the same hardware specification and the same connection speed, so the impact of the slowest clients on the overall duration mostly depends on the size of its local training set and the computational complexity of the local functions, including the loss function, model aggregation and evaluation. Another factor that can influence the duration is the amount of data exchanged between the clients and the server, which influences the transmission time. FLAD, DAFL and FedSBS always send the global model to all clients (either for training or evaluation), while \ac{fvg}, FedProx and SCAFFOLD only send the model to the selected clients. In addition, FedSBS sends local evaluation results to the server, like FLAD, which also distributes the training epochs and steps assigned to each client. These additional interactions contribute to the overall duration of the \ac{fl} process.

\subsection{Local execution time}\label{sec:results:local-time}
As mentioned above, the local execution time of each client influences the overall duration of the \ac{fl} process. This time determines how quickly each round can be completed, as the server must wait for all selected clients to finish their local operations before proceeding to the next round. In addition, this value also indicates, along with the number of rounds a client is selected for training, the total computational load imposed on each client.

\begin{figure*}[t!]
	\begin{subfigure}[t]{0.49\textwidth} 
            \begin{tikzpicture}
    \pgfplotstableread[col sep=comma]{results_data/fedavg_time_per_client_new.dat}\datatable
    \pgfplotsset{boxplot/whisker range=1000}

    \begin{axis}[
        title={\ac{fvg}},
        width=\textwidth,height=5cm,
        scaled y ticks=base 10:-3,
        ylabel={Time ($\times 10^3$ seconds)},
        scaled y ticks=false,
        ymin=0,
        ymax=18000,
        ytick distance=2000,
        ytick={0,2000,...,18000},
        yticklabels={0,2,...,18},
        grid=both,
        boxplot/draw direction=y,
        xtick={1,...,15},
        xticklabels={WebDDoS,LDAP,Portmap,DNS,UDPLag,NTP,SNMP,SSDP,Syn,TFTP,UDP,NetBIOS,MSSQL,LOIC,HOIC},
        % rotate labels by 45 degres
        xticklabel style={rotate=40, anchor=east},
        % style for the box bodies
        boxplot/every box/.style={
        draw=black,            % box outline color
        fill=white!30,          % box fill color (same for all)
        line width=0.8pt
        },
        % style for the median line
        boxplot/every median/.style={
        draw=blue!90!black,
        line width=1.2pt
        },
        % style for whiskers and whisker caps
        boxplot/every whisker/.style={
        draw=black,
        line width=0.9pt
        },
        boxplot/every whisker cap/.style={
        draw=black,
        line width=0.9pt
        },
        % reduce horizontal gap between boxes
        boxplot/box extend=0.5,
        enlargelimits=0.05
        ]
        % Add one boxplot per column. The loop uses the column name as y=...
        \foreach \col in {WebDDoS,LDAP,Portmap,DNS,UDPLag,NTP,SNMP,SSDP,Syn,TFTP,UDP,NetBIOS,MSSQL,LOIC,HOIC} {
        \addplot[boxplot] table[y=\col] {\datatable};
        }

    \end{axis}
    \end{tikzpicture}
	\end{subfigure}  
    \begin{subfigure}[t]{0.49\textwidth} 
            \begin{tikzpicture}
    \pgfplotstableread[col sep=comma]{results_data/dafl_time_per_client.dat}\datatable
    \pgfplotsset{boxplot/whisker range=1000}

    \begin{axis}[
        title={DAFL},
        width=\textwidth,height=5cm,
        scaled y ticks=base 10:-3,
        ylabel={Time ($\times 10^3$ seconds)},
        scaled y ticks=false,
        ymin=0,
        ymax=18000,
        ytick distance=2000,
        ytick={0,2000,...,18000},
        yticklabels={0,2,...,18},
        grid=both,
        boxplot/draw direction=y,
        xtick={1,...,15},
        xticklabels={WebDDoS,LDAP,Portmap,DNS,UDPLag,NTP,SNMP,SSDP,Syn,TFTP,UDP,NetBIOS,MSSQL,LOIC,HOIC},
        % rotate labels by 45 degres
        xticklabel style={rotate=40, anchor=east},
        % style for the box bodies
        boxplot/every box/.style={
        draw=black,            % box outline color
        fill=white!30,          % box fill color (same for all)
        line width=0.8pt
        },
        % style for the median line
        boxplot/every median/.style={
        draw=blue!90!black,
        line width=1.2pt
        },
        % style for whiskers and whisker caps
        boxplot/every whisker/.style={
        draw=black,
        line width=0.9pt
        },
        boxplot/every whisker cap/.style={
        draw=black,
        line width=0.9pt
        },
        % reduce horizontal gap between boxes
        boxplot/box extend=0.5,
        enlargelimits=0.05
        ]
        % Add one boxplot per column. The loop uses the column name as y=...
        \foreach \col in {WebDDoS,LDAP,Portmap,DNS,UDPLag,NTP,SNMP,SSDP,Syn,TFTP,UDP,NetBIOS,MSSQL,LOIC,HOIC} {
        \addplot[boxplot] table[y=\col] {\datatable};
        }

    \end{axis}
    \end{tikzpicture}
	\end{subfigure}
	\caption{Distribution of local training times for each client with \ac{fvg} and DAFL ($10^3$ scale). The boxplots indicate the interquartile range (i.e., the range between the 25th and 75th percentiles), with the horizontal blue line denoting the average value.}
	\label{fig:fedavg-others-duration}
\end{figure*}

Figure \ref{fig:fedavg-others-duration} shows the distribution of local training times of each client for \ac{fvg}, and DAFL. FedProx, SCAFFOLD, FedALA and FedSBS are not included in this analysis, as they show similar results to \ac{fvg} (see Figure \ref{fig:local_duration}).
The figure demonstrates the dependency of the local training time on the size of the local dataset, which is consistent across all methods. The clients with larger datasets, such as \textit{MSSQL} and \textit{HOIC}, require more time to train their local models compared to others. However, while with \ac{fvg} (as well as FedProx, SCAFFOLD, FedALA and FedSBS) such clients are not always selected, with DAFL they are selected in every round, leading to a longer overall duration of the \ac{fl} process.
% In addition, the figure shows the impact of the additional computations required by some methods. For example, FedALA requires an extra training phase to estimate the aggregation weights, while FedSBS requires an evaluation phase, which also adds to the local training time.

\begin{figure}[t!]
    \begin{tikzpicture}
    \pgfplotstableread[col sep=comma]{results_data/flad_time_per_client.dat}\datatable
    \pgfplotsset{boxplot/whisker range=1000}

    \begin{axis}[
        title={FLAD},
        width=0.49\textwidth,height=5cm,
        scaled y ticks=base 10:-2,
        ylabel={Time ($\times 10^2$ seconds)},
        scaled y ticks=false,
        ymin=0,
        ymax=600,
        ytick distance=100,
        ytick={0,100,...,600},
        yticklabels={0,1,...,6},
        grid=both,
        boxplot/draw direction=y,
        xtick={1,...,15},
        xticklabels={WebDDoS,LDAP,Portmap,DNS,UDPLag,NTP,SNMP,SSDP,Syn,TFTP,UDP,NetBIOS,MSSQL,LOIC,HOIC},
        % rotate labels by 45 degres
        xticklabel style={rotate=40, anchor=east},
        % style for the box bodies
        boxplot/every box/.style={
        draw=black,            % box outline color
        fill=white!30,          % box fill color (same for all)
        line width=0.8pt
        },
        % style for the median line
        boxplot/every median/.style={
        draw=blue!90!black,
        line width=1.2pt
        },
        % style for whiskers and whisker caps
        boxplot/every whisker/.style={
        draw=black,
        line width=0.9pt
        },
        boxplot/every whisker cap/.style={
        draw=black,
        line width=0.9pt
        },
        % reduce horizontal gap between boxes
        boxplot/box extend=0.5,
        enlargelimits=0.05
        ]
        % Add one boxplot per column. The loop uses the column name as y=...
        \foreach \col in {WebDDoS,LDAP,Portmap,DNS,UDPLag,NTP,SNMP,SSDP,Syn,TFTP,UDP,NetBIOS,MSSQL,LOIC,HOIC} {
        \addplot[boxplot] table[y=\col] {\datatable};
        }
    \end{axis}
    \end{tikzpicture}
	\caption{Distribution of local training times for each client with FLAD ($10^2$ scale). The boxplots indicate the interquartile range (i.e., the range between the 25th and 75th percentiles), with the horizontal blue line denoting the average.}
	\label{fig:fedavg-flad-duration}
\end{figure}

A different behaviour is observed with FLAD, as shown in Figure \ref{fig:fedavg-flad-duration} (please note that, in this figure, the y-axis is scaled by a factor of $10^2$ to improve readability). The local training time of each client is influenced not only by the size of its local dataset, but also by the number of training rounds, epochs, and steps assigned to it, which are adaptively determined based on the performance of the global model on the client's local validation set. As a result, clients with smaller datasets and harder-to-learn attacks, such as \textit{WebDDoS}, \textit{LDAP}, and \textit{Portmap}, may require a total training time comparable to, or even longer than, that of clients with larger datasets, such as \textit{UDP}, \textit{NetBIOS}, and \textit{MSSQL}.
The adaptive approach of FLAD can also be observed by comparing the considerably longer local training time of \textit{LOIC} with that of \textit{TFTP}, despite their local datasets being of similar size. This difference can be attributed to the more challenging attack represented by \textit{LOIC}.
\subsection{Aggregation analysis}\label{sec:results:aggregation}

% histogram of the validation accuracy of the global model and of the local models before aggregation
\begin{figure*}[!t]
\begin{subfigure}[t]{0.49\textwidth} 
\begin{tikzpicture}
\begin{axis}[
    ybar,
    bar width=3pt,
    width=1.09\linewidth,
    height=4.5cm,
    ylabel={F1 Score},
    symbolic x coords={FedAvg,FedProx,SCAFFOLD,FedALA,DAFL,FedSBS,FLAD},
    xtick=data,
    enlarge x limits=0.1,
    xtick style={draw=none},
    xticklabel style={rotate=40, anchor=east},
    % nodes near coords,
    % nodes near coords align={vertical},
    legend style={at={(1.05,1.05)},anchor=south,legend columns=-1},
    ymin=0, ymax=1.05,
    ytick={0,0.2,0.4,0.6,0.8,1.0},
    yticklabels={0,0.2,0.4,0.6,0.8,1.0},
    grid=both,
    every node near coord/.append style={text=black}
]

% Define CSV content once
\pgfplotstableread[col sep=comma]{results_data/aggregation_local.dat}\datatable

% Plot WebDDoS
\addplot+[
    pattern=crosshatch, pattern color=cb_blue, draw=cb_blue
] table[x=Method,y=WebDDoS]{\datatable};

% Plot Syn
\addplot+[
    pattern=north east lines, pattern color=cb_aquamarine, draw=cb_aquamarine
] table[x=Method,y=LDAP]{\datatable};

% Plot NetBIOS
\addplot+[
    pattern=crosshatch dots
    , pattern color=cb_caramel, draw=cb_caramel
] table[x=Method,y=NetBIOS]{\datatable};

% Plot MSSQL
\addplot+[
    pattern=north west lines
    , pattern color=gray, draw=gray
] table[x=Method,y=MSSQL]{\datatable};

% Plot LOIC
\addplot+[
    style={cb_red,fill=cb_red!40}
] table[x=Method,y=LOIC]{\datatable};

\legend{WebDDoS,LDAP,NetBIOS,MSSQL,LOIC}

\end{axis}
\end{tikzpicture}
\vspace{-9mm}
\caption{Local model.}
\label{fig:local-global:local}
\end{subfigure}
\hspace{3mm}
\begin{subfigure}[t]{0.49\textwidth} 
\begin{tikzpicture}
\begin{axis}[
    ybar,
    bar width=3pt,
    width=1.09\linewidth,
    height=4.5cm,
    symbolic x coords={FedAvg,FedProx,SCAFFOLD,FedALA,DAFL,FedSBS,FLAD},
    xtick=data,
    enlarge x limits=0.1,
    xtick style={draw=none},
    xticklabel style={rotate=40, anchor=east},
    % nodes near coords,
    % nodes near coords align={vertical},
    ymin=0, ymax=1.05,
    ytick={0,0.2,0.4,0.6,0.8,1.0},
    yticklabels={0,0.2,0.4,0.6,0.8,1.0},
    grid=both,
    every node near coord/.append style={text=black}
]

% Define CSV content once
\pgfplotstableread[col sep=comma]{results_data/aggregation_global.dat}\datatable

% Plot WebDDoS
\addplot+[
    pattern=crosshatch, pattern color=cb_blue, draw=cb_blue
] table[x=Method,y=WebDDoS]{\datatable};

% Plot Syn
\addplot+[
    pattern=north east lines, pattern color=cb_aquamarine, draw=cb_aquamarine
] table[x=Method,y=LDAP]{\datatable};

% Plot NetBIOS
\addplot+[
    pattern=crosshatch dots
    , pattern color=cb_caramel, draw=cb_caramel
] table[x=Method,y=NetBIOS]{\datatable};

% Plot MSSQL
\addplot+[
    pattern=north west lines
    , pattern color=gray, draw=gray
] table[x=Method,y=MSSQL]{\datatable};

% Plot LOIC
\addplot+[
    style={cb_red,fill=cb_red!40}
] table[x=Method,y=LOIC]{\datatable};

\end{axis}
\end{tikzpicture}
\vspace{-5mm}
\caption{Global model.}
\label{fig:local-global:global}
\end{subfigure}
\caption{F1 scores on the validation datasets of the local models (left) and the global model (right) after aggregation. Only the clients that show significant changes in accuracy after aggregation are included.}
    \label{fig:local-global}
\end{figure*}

In this section, we analyse the performance of the aggregation mechanisms implemented by the different methods. We focus on the validation accuracy of the global model and of the local models before aggregation, as well as on the impact of aggregation on these scores. This analysis allows us to understand how the different aggregation and client selection mechanisms (see Table \ref{tab:algorithms}) affect the accuracy of the global model in correctly classifying the attacks. 

The plots in Figure \ref{fig:local-global} compare the performance of the local models with that of the aggregated global model. The reported scores correspond to the training round in which the global model achieves its highest average performance. To ensure a fair assessment of the aggregation mechanisms, we evaluate the local models that contributed to the derivation of that global model. Moreover, we focus on a subset of representative clients, chosen to better highlight the impact of the aggregation mechanisms on the global model's performance.
Performance is evaluated using the F1 score, the harmonic mean of precision (P) and recall (R), which are formally defined as follows:
$$F1=2\cdot\frac{P\cdot R}{P + R}\quad P=\frac{TP}{TP+FP}\quad R=\frac{TP}{TP+FN}$$

\noindent where \textit{TP=True Positives}, \textit{FP=False Positives} and \textit{FN=False Negatives}.

% As expected, local models generally perform better on their respective datasets.
% However, aggregation can significantly impact this performance.
% Clients with smaller datasets, such as those associated with the WebDDoS and LDAP attacks, or with \ac{ood} attacks, like WebDDos and LOIC, show a notable drop in accuracy after aggregation, while large UDP-based ones, like \textit{NetBIOS} and \textit{MSSQL}, are less affected by this process.

\begin{figure*}[b!]
\begin{subfigure}[t]{0.49\textwidth} 
\pgfplotstableread[col sep=comma]{results_data/webddos_validation_f1.dat}{\accuracy}
    \centering
    \begin{tikzpicture}
        \begin{axis}[
            legend columns = 7,
            legend style = {at={(1, 1.02)}, anchor=south, style={column sep=0.15cm}},       %vorher: at={(1.3, 1)}
            legend cell align=left,
            xmin = 0, xmax=250, %4100
            ymin = 0, ymax = 1, %155000
            xtick distance=25,
            xtick={0,25,...,250},
            xticklabels={0,25,...,250},
            xtick pos=left,
            ytick distance=0.1,
            ytick={0.1,0.2,0.3,0.4,0.5,0.6,0.7,0.8,0.9,1},
            yticklabels={0.1,0.2,0.3,0.4,0.5,0.6,0.7,0.8,0.9,1},
            grid = both,
            width=1.05\linewidth,
            height=5cm,
            xlabel = {Round},
            xlabel style={
                at={(ticklabel* cs:0)}, % position at the left xtick
                anchor=north east       % align above the tick
            },  
            ylabel = {F1 Score},
            enlargelimits=0.02
            ]
    
            % Layers DAFL FedALA FedAvg FedProx FedSBS FLAD SCAFFOLD
            \addplot[color=fvg, style={thick},mark=diamond*] table [x = {Epoch}, y = {FedAVG}] {\accuracy};
            \addplot[color=fedprox, style={thick}, mark=square*, mark size=1.5] table [x = {Epoch}, y = {FedProx}] {\accuracy};
            \addplot[color=scaffold, style={thick},mark=triangle*] table [x = {Epoch}, y = {SCAFFOLD}] {\accuracy};
            \addplot[color=fedala,style={thick}, mark=triangle] table [x = {Epoch}, y = {FedALA}] {\accuracy};
            \addplot[color=dafl, style={thick}, mark=square, mark size=1.5] table [x = {Epoch}, y = {DAFL}] {\accuracy};
            \addplot[color=fedsbs, style={thick}, mark=o,mark size=1.5] table [x = {Epoch}, y = {FedSBS}] {\accuracy};
            \addplot[color=flad, style={thick},mark=*, mark size=1.5] table [x = {Epoch}, y = {FLAD}] {\accuracy};

            \legend{\ac{fvg},FedProx,SCAFFOLD,FedALA,DAFL,FedSBS,FLAD}

            %\legend{FedAvg,FedProx,SCAFFOLD,FedALA,DAFL,FedSBS,FLAD}
        \end{axis}
    \end{tikzpicture}
    \captionsetup{skip=2pt}
    \caption{Validation F1 score of the WebDDoS client.}
    \label{fig:f1-validation-webddos}
    \vspace{2mm}
\end{subfigure}
\begin{subfigure}[t]{0.49\textwidth} 
\pgfplotstableread[col sep=comma]{results_data/loic_validation_f1.dat}{\accuracy}
    \centering
    \begin{tikzpicture}
        \begin{axis}[
            % legend on two rows
            legend columns = 2,
            legend style = {at={(0.45, 0.05)}, anchor=south west, style={column sep=0.15cm}},       %vorher: at={(1.3, 1)}
            legend cell align=left,
            xmin = 0, xmax=250, %4100
            ymin = 0, ymax = 1, %155000
            xtick distance=25,
            xtick={0,25,...,250},
            xticklabels={0,25,...,250},
            xtick pos=left,
            ytick distance=0.1,
            ytick={0.1,0.2,0.3,0.4,0.5,0.6,0.7,0.8,0.9,1},
            yticklabels={0.1,0.2,0.3,0.4,0.5,0.6,0.7,0.8,0.9,1},
            grid = both,
            width=1.05\linewidth,
            height=5cm,
            xlabel = {Round},
            xlabel style={
                at={(ticklabel* cs:0)}, % position at the left xtick
                anchor=north east       % align above the tick
            },  
            ylabel = {F1 Score},
            enlargelimits=0.02
            ]
    
            % Layers DAFL FedALA FedAvg FedProx FedSBS FLAD SCAFFOLD
            \addplot[color=fvg, style={thick},mark=diamond*] table [x = {Epoch}, y = {FedAVG}] {\accuracy};
            \addplot[color=fedprox, style={thick}, mark=square*, mark size=1.5] table [x = {Epoch}, y = {FedProx}] {\accuracy};
            \addplot[color=scaffold, style={thick},mark=triangle*] table [x = {Epoch}, y = {SCAFFOLD}] {\accuracy};
            \addplot[color=fedala,style={thick}, mark=triangle] table [x = {Epoch}, y = {FedALA}] {\accuracy};
            \addplot[color=dafl, style={thick}, mark=square, mark size=1.5] table [x = {Epoch}, y = {DAFL}] {\accuracy};
            \addplot[color=fedsbs, style={thick}, mark=o,mark size=1.5] table [x = {Epoch}, y = {FedSBS}] {\accuracy};
            \addplot[color=flad, style={thick},mark=*, mark size=1.5] table [x = {Epoch}, y = {FLAD}] {\accuracy};
        \end{axis}
    \end{tikzpicture}
    \captionsetup{skip=2pt}
    \caption{Validation F1 score of the LOIC client.}
    \label{fig:f1-validation-loic}
\end{subfigure}
\caption{F1 score of the global model over increasing rounds, evaluated on the validation sets of the WebDDoS and LOIC clients. Results are from the longest experiment (250 rounds).}
    \label{fig:f1-validation-ood}
\end{figure*}

\ac{fvg}, FedProx, SCAFFOLD, DAFL and FedSBS show comparable performance on the WebDDoS, LDAP and LOIC attacks. While they achieve satisfactory accuracy before aggregation on all the three attacks, after aggregation they struggle to maintain a similar level of accuracy. These findings are consistent with previous research \cite{flad}, which highlighted the limitations of relying on random client selection and weighted averaging as aggregation strategies in such scenarios (though FedSBS does not rely on purely random client selection). DAFL in contrast, implements a client selection strategy that prioritises clients with higher validation accuracy. However, this approach does not produce satisfying results of the global model on these three attacks. 

FedALA aggregates each client's local model with the global model received from the server, aiming to adapt the global model to the client's local objective. Although this approach improves performance on local datasets compared to \ac{fvg} (see Figure \ref{fig:local-global:local}), once the server aggregates these adapted models, the resulting global model loses its ability to generalise to two of the four TCP-based attacks, namely \textit{WebDDoS} and \textit{LOIC} (see Figure \ref{fig:local-global:global}).

FLAD shows high robustness to aggregation, producing a global model capable of effectively classifying all attacks. This is achieved by leveraging the feedback from clients regarding the performance of the global model on their local datasets, enabling the server to dynamically adjust the training process and concentrate on the most challenging attacks. 

The limitations \ac{fvg} and the other methods in handling some of the attacks are evident in Figure \ref{fig:f1-validation-ood}, which report the F1 score of the aggregated global model on the WebDDoS and LOIC attacks across training rounds. The plots indicate that, with the exception of FLAD, all methods struggle to generate a global model capable of effectively classifying these attacks. The WebDDoS attack, in particular, proves especially challenging, likely due to the relatively small size of its dataset compared to the others.
This limits aggregation mechanisms that weight contributions by dataset size, causing clients with smaller datasets to be underrepresented in the global model. 

% Figures \ref{fig:f1-validation-overall} illustrates the average validation F1 score of the global model over increasing rounds. The plot shows that, except for FedALA, all the methods achieve a satisfactory average F1 score ($0.9$ or above) at the end of the training process.

A key observation in this context is the number of rounds required to reach a given F1 score threshold, as fewer rounds allow clients to receive the final updated global model for inference sooner. Figure \ref{fig:f1-threshold} shows that FLAD and \ac{fvg} are the fastest methods to reach robust F1 scores of $0.8$ and $0.9$, while FedProx and FedSBS are the slowest, requiring more than $70$ rounds to reach an F1 score of $0.9$. FedALA reaches an F1 score of $0.8$ in around $45$ rounds, but fails to reach $0.9$ altogether.

\begin{figure}[h!]
\centering
\begin{tikzpicture}
\begin{axis}[
    legend columns = 2,
    legend style = {at={(0.5, 1.02)}, anchor=south, style={column sep=0.15cm}},       %vorher: at={(1.3, 1)}
    legend cell align=left,
    legend pos=south east,
    xlabel = {Round},
    xlabel style={
        at={(ticklabel* cs:0)}, % position at the left xtick
        anchor=north east       % align above the tick
    },  
    ylabel={F1 Score Threshold},
    grid=major,
    width=1.05\linewidth,
    height=6cm,
    xmin = 0, xmax=90, %4100
    ymin = 0.5, ymax = 1, %155000
    xtick distance=10,
    xtick={0,10,...,90},
    xticklabels={0,10,...,90},
    ytick={0.5,0.6,0.7,0.8,0.9,1},
    yticklabels={0.5,0.6,0.7,0.8,0.9,1},
    enlargelimits=0.02
]
\addplot[color=fvg, style={thick},mark=diamond*] coordinates {
    (2.90,0.5)
    (4.70,0.6)
    (5.70,0.7)
    (8.40,0.8)
    (20.60,0.9)
};
\addlegendentry{\ac{fvg}}
\addplot[color=fedprox, style={thick}, mark=square*, mark size=1.5] coordinates {
    (2.50,0.5)
    (5.10,0.6)
    (16.90,0.7)
    (44.40,0.8)
    (70.80,0.9)
};
\addlegendentry{FedProx}
\addplot[color=scaffold, style={thick}, mark=triangle*] coordinates {
    (1.40,0.5)
    (2.50,0.6)
    (5.50,0.7)
    (10.10,0.8)
    (54.44,0.9)
};
\addlegendentry{SCAFFOLD}
\addplot[color=fedala, style={thick}, mark=triangle] coordinates {
    (1.50,0.5)
    (2.70,0.6)
    (10.70,0.7)
    (44.67,0.8)
};
\addlegendentry{FedALA}
\addplot[color=dafl, style={thick}, mark=square, mark size=1.5] coordinates {
    (2.00,0.5)
    (2.00,0.6)
    (5.50,0.7)
    (10.60,0.8)
    (54.10,0.9)
};
\addlegendentry{DAFL}
\addplot[color=fedsbs, style={thick}, mark=o, mark size=1.5] coordinates {
    (2.40,0.5)
    (3.90,0.6)
    (9.90,0.7)
    (19.90,0.8)
    (84.71,0.9)
};
\addlegendentry{FedSBS}
\addplot[color=flad, style={thick}, mark=o, mark size=1.5pt] coordinates {
    (2.30,0.5)
    (3.40,0.6)
    (4.60,0.7)
    (5.80,0.8)
    (18.10,0.9)
};
\addlegendentry{FLAD}

\end{axis}
\end{tikzpicture}

\captionsetup{skip=2pt}
\caption{Rounds taken to reach a certain F1 score threshold.}
\label{fig:f1-threshold}
\end{figure}

\begin{table*}[b!]
    \caption{Test set accuracy (F1 score). The values in bold indicate the scores that are significantly lower than $0.80$.}
    \label{tab:test-f1}
    \centering
    \begin{threeparttable}
    \resizebox{\textwidth}{!}{%
    \begin{tabular}{lccccccccccccccc|cc}
    \toprule[\heavyrulewidth]
    \textbf{Method} & \textbf{WebDDoS} & \textbf{LDAP} & \textbf{Portmap} & \textbf{DNS} & \textbf{UDPLag} & \textbf{NTP} & \textbf{SNMP} & \textbf{SSDP} & \textbf{Syn} & \textbf{TFTP} & \textbf{UDP} & \textbf{NetBIOS} & \textbf{MSSQL} & \textbf{LOIC} & \textbf{HOIC} & \textbf{Mean} & \textbf{Std\_dev} \\ 
    \midrule[\heavyrulewidth]
    \textbf{\ac{fvg}} 
        & \textbf{0.78} & \textbf{0.61} & 0.88 & 0.92 & 0.90 & 0.93 & 0.95 & 0.98 & 0.91 & 0.99 & 1.00 & 0.96 & 0.99 & \textbf{0.56} & 1.00 & 0.889 & 0.137 \\
    \textbf{FedProx} 
        & \textbf{0.02} & \textbf{0.77} & 0.89 & 0.90 & 0.91 & 0.89 & 0.99 & 0.99 & 0.85 & 0.99 & 1.00 & 0.97 & 1.00 & \textbf{0.68} & 0.99 & 0.856 & 0.248 \\
    \textbf{SCAFFOLD} 
        & \textbf{0.62} & \textbf{0.54} & 0.89 & 0.86 & 0.90 & 0.91 & 0.95 & 0.99 & 0.91 & 0.99 & 1.00 & 0.96 & 0.99 & \textbf{0.57} & 1.00 & 0.871 & 0.160 \\
    \textbf{FedALA} 
        & \textbf{0.36} & \textbf{0.46} & \textbf{0.48} & \textbf{0.42} & \textbf{0.60} & 0.89 & \textbf{0.79} & 0.92 & \textbf{0.11} & 0.85 & 0.95 & \textbf{0.49} & \textbf{0.69} & \textbf{0.01} & \textbf{0.40} & 0.562 & 0.287 \\
    \textbf{DAFL} 
        & \textbf{0.77} & \textbf{0.66} & 0.88 & 0.95 & 0.93 & 0.93 & 0.95 & 0.99 & 0.91 & 0.99 & 1.00 & 0.94 & 0.99 & \textbf{0.67} & 1.00 & 0.904 & 0.113 \\
    \textbf{FedSBS} 
        & \textbf{0.07} & \textbf{0.65} & 0.92 & 0.94 & 0.97 & 0.90 & 0.99 & 0.99 & 1.00 & 1.00 & 1.00 & 0.98 & 1.00 & \textbf{0.73} & 1.00 & 0.877 & 0.245 \\
    \textbf{FLAD} 
        & 0.97 & 0.96 & 0.95 & 0.98 & 0.98 & 0.98 & 0.98 & 1.00 & 0.99 & 0.99 & 0.98 & 0.98 & 0.98 & 0.97 & 0.98 & 0.978 & 0.012 \\
    \bottomrule[\heavyrulewidth]
    \end{tabular}}
    \end{threeparttable}
\end{table*}

\subsection{Test accuracy}
In the previous section, we evaluated the performance of the global models on the clients' validation sets to highlight the impact of different aggregation and client selection strategies on the learning process. We now assess the final global models on the clients' test sets, simulating the deployment scenario in which the global model is distributed to all clients and applied to unseen data. As in the training and validation phases, we assume that each client’s test set contains only samples of a single attack type (along with benign traffic).

The results reported in Table \ref{tab:test-f1} are obtained by evaluating each of the 10 global models trained in the previous experiments on each client’s test set. For each client, the resulting F1 scores are first averaged across the 10 experiments. The last two columns then report the mean and standard deviation of these per-client average F1 scores across all 15 clients, providing an overall measure of the global model's performance. These results largely confirm the trend observed during the \ac{fl} process and described in Section \ref{sec:results:aggregation}.

Overall, we observe that most attacks are correctly identified, F1 score over 0.9, by all methods, with the exception of \textit{WebDDoS}, \textit{LDAP}, and \textit{LOIC}, which present significant challenges for all methods. In particular, the \textit{WebDDoS} and \textit{LDAP} attacks are the most difficult to detect, with F1 scores below $0.8$ for all methods except FLAD. This is likely due to the fact that these attacks are underrepresented in the training data, making it difficult for the global model to learn their characteristics due to weighted aggregation and client selection mechanisms that favour clients with more training samples.
\textit{LOIC} is also challenging for most methods, with F1 scores below $0.8$, except for FLAD. Although the training set size is relatively large for this attack, it is a TCP-based attack, whose feature distribution is significantly different from the other attacks, which are mostly UDP-based. 

Notably, DAFL emerges as the second-best method after FLAD, achieving performance comparable to \ac{fvg} in terms of average F1 score and standard deviation. However, it incurs higher computational costs due to the fact that all clients must train in each round. Other approaches, such as FedProx, SCAFFOLD, FedALA and FedSBS, yield lower average F1 scores and underperform on at least two attack types.

%In summary, although all the methods are designed to address the client-drift issues of \ac{fvg}, only the adaptive mechanism of FLAD succeeds in learning effectively from all the datasets. 
%This suggests that requiring additional information from clients, such as validation accuracy scores, is essential for enhancing client selection and optimising local training, thereby addressing challenges posed by heterogeneous data distributions in \ac{fl}.  
\section{Discussion and conclusion}\label{sec:conclusions}

This study evaluates the performance of state-of-the-art \ac{fl} algorithms for cybersecurity, with a focus on network intrusion detection where \ac{ml} models are trained on network traffic. We consider a representative set of FL algorithms addressing the challenges posed by non-\ac{iid} and unbalanced datasets. These algorithms differ in how they address client heterogeneity, including modifications to local training, model aggregation and client selection.

The objective of this work is to understand how different \ac{fl} training algorithms behave in cybersecurity scenarios characterised by challenging client-level heterogeneity and to quantify the resulting trade-offs between detection performance, communication overhead, and training time.

Overall, the results suggest that how clients are selected and how training workloads are distributed across them can be particularly important in strongly non-\ac{iid} scenarios. Algorithms that explicitly account for differences between clients and adapt their participation or training workload accordingly achieve higher detection performance than approaches that primarily modify the local optimisation or global aggregation process while maintaining a fixed client participation scheme. In our setting, the most effective algorithms (DAFL and FLAD) dynamically prioritise clients according to their contribution to the global learning process.% and allocate more training resources to clients whose local performance indicates that their data are not yet adequately represented by the global model.

This observation is particularly relevant for network intrusion detection. Under highly heterogeneous data distributions, different clients may observe substantially different attack types, with some attacks being available only at a small subset of clients. Treating all clients equally, either by selecting them with a fixed probability or by assigning them the same training workload, does not necessarily ensure that all attack types contribute sufficiently to the global model. Conversely, an informed selection mechanism can focus the training process on clients that provide information currently underrepresented in the global model.

%The network overhead associated with such adaptive strategies is higher than that of conventional FL approaches with fixed client participation. This overhead results from the additional information required to assess client performance and determine which clients should participate in subsequent rounds. In our setting, this is not a critical limitation, as the model size is small and the communication infrastructure is efficient. However, in scenarios involving larger models or constrained bandwidth, the additional communication required by adaptive client selection could become a relevant bottleneck.

The overall duration of the \ac{fl} process is also influenced by how client participation and training workloads are managed. Fixed participation schemes may repeatedly involve computationally expensive clients, even when their contribution is not critical to the current training stage. In contrast, adaptive algorithms can reduce training time by excluding clients that provide limited additional benefit. This is particularly relevant in cybersecurity scenarios, where timely updates of the global model are essential to enable a prompt response to new or emerging attack types.

%These results suggest that, under severe client-level heterogeneity, adapting the distribution of the training workload may be more effective than relying exclusively on improvements to local training or model aggregation. Local optimisation and aggregation strategies can mitigate some effects of statistical heterogeneity, but they do not directly address the question of which clients should contribute to each global update or how much computational effort should be devoted to each client. When different clients represent substantially different portions of the attack space, these decisions become central to ensuring that the global model adequately incorporates the information available across the federation.

%This observation is particularly important in dynamic cybersecurity environments. The attacks encountered by individual clients cannot generally be known a priori, and clients may become exposed to new or previously unseen attack types over time. An effective federated training strategy should therefore be able to adapt to changes in the information available across clients, ensuring that rare or locally concentrated attack types are not systematically underrepresented in the global model.

From this perspective, the main finding of our experiments is that informed client selection and adaptive workload distribution provide a particularly effective mechanism for addressing severe non-\ac{iid} and unbalanced data distributions. By dynamically determining which clients should participate and how much training effort should be allocated to them, the federation can make more effective use of the heterogeneous information available across clients. This comes at the cost of additional communication and coordination, highlighting a trade-off between adaptivity and communication efficiency. Nevertheless, the results indicate that, for cybersecurity scenarios characterised by highly heterogeneous attack distributions, investing in informed participation and workload allocation can provide greater benefits than modifying local training or aggregation strategies alone.

\section*{Acknowledgment}
This work was supported by Ministero delle Imprese e del Made in Italy (IPCEI Cloud DM 27 giugno 2022 - IPCEI-CL-0000007) and European Union (Next Generation EU).

\bibliographystyle{elsarticle-num}
\bibliography{bibliography}

\end{document}